\documentclass[epj,nopacs]{svjour}
\usepackage{graphics}
\usepackage{graphicx}
\usepackage{color}

\newcommand{\la}{\left<}
\newcommand{\ra}{\right>}

\newcommand{\nvecl}{\underline{n}_l}

\newcommand{\rvec}{\ensuremath{\underline{r}}}
\newcommand{\rvecl}{\ensuremath{\underline{r}_l}}

\newcommand{\ddiff}{\ensuremath{\mathrm{d}}}

\newcommand{\rl}{r_{l}}

\newcommand{\nlx}{n_{l,x}}
\newcommand{\nly}{n_{l,y}}

\newcommand{\kB}{k_\mathrm{B}}

\newcommand{\tauhat}{\ensuremath{\hat{\sigma}}}
\newcommand{\muA}{\ensuremath{\mu_\mathrm{A}}}

\newcommand{\muF}{\ensuremath{\mu_\mathrm{F}}}

\newcommand{\tincr}{\delta t}
\newcommand{\tsamp}{\Delta t}
\newcommand{\tsampmax}{{\Delta t}_{\mathrm{max}}}
\newcommand{\tspacer}{{\Delta t}_{\mathrm{spac}}}
\newcommand{\ttemper}{{\Delta t}_{\mathrm{temp}}}

\newcommand{\xbf}{\ensuremath{\mathbf{x}}}

\newcommand{\Nc}{\ensuremath{n_\mathrm{c}}}

\newcommand{\Nk}{\ensuremath{n_\mathrm{k}}}
\newcommand{\Nl}{\ensuremath{n_l}}
\newcommand{\Nt}{\ensuremath{n_\mathrm{t}}}
\newcommand{\Nm}{\ensuremath{n_\mathrm{m}}}

\newcommand{\Ocal}{\ensuremath{{\cal O}}}

\newcommand{\Aop}{\ensuremath{\mathbf{A}}}
\newcommand{\Bop}{\ensuremath{\mathbf{B}}}

\newcommand{\Eop}{\ensuremath{\mathbf{E}}}

\newcommand{\Pop}{\ensuremath{\mathbf{P}}}
\newcommand{\Vop}{\ensuremath{\mathbf{V}}}

\newcommand{\dOtot}{\ensuremath{\delta \Ocal^2_{\mathrm{tot}}}}
\newcommand{\dOint}{\ensuremath{\delta \Ocal^2_{\mathrm{int}}}}
\newcommand{\dOext}{\ensuremath{\delta \Ocal^2_{\mathrm{ext}}}}
\newcommand{\sOtot}{\ensuremath{\delta \Ocal_{\mathrm{tot}}}}
\newcommand{\sOint}{\ensuremath{\delta \Ocal_{\mathrm{int}}}}
\newcommand{\sOext}{\ensuremath{\delta \Ocal_{\mathrm{ext}}}}
\newcommand{\dvtot}{\ensuremath{\delta v^2_{\mathrm{tot}}}}
\newcommand{\svtot}{\ensuremath{\delta v_{\mathrm{tot}}}}
\newcommand{\dvint}{\ensuremath{\delta v^2_{\mathrm{int}}}}
\newcommand{\svint}{\ensuremath{\delta v_{\mathrm{int}}}}
\newcommand{\dvext}{\ensuremath{\delta v^2_{\mathrm{ext}}}}
\newcommand{\svext}{\ensuremath{\delta v_{\mathrm{ext}}}}
\newcommand{\dvgauss}{\ensuremath{\delta v^2_{\mathrm{G}}}}
\newcommand{\svgauss}{\ensuremath{\delta v_{\mathrm{G}}}}

\newcommand{\sBptot}{\ensuremath{\delta B_{p,\mathrm{tot}}}}
\newcommand{\sBpint}{\ensuremath{\delta B_{p,\mathrm{int}}}}

\newcommand{\sinf}{\ensuremath{s_{\infty}}}

\newcommand{\taualph}{\tau_{\alpha}}

\newcommand{\Dnonerg}{\Delta_{\mathrm{ne}}^2}
\newcommand{\Snonerg}{\Delta_{\mathrm{ne}}}
\newcommand{\Tnonerg}{\tau_{\mathrm{ne}}}

\newcommand{\Tglass}{T_{\mathrm{g}}}

\newcommand{\taualpha}{\tau_{\alpha}}

\newcommand{\taubasin}{\tau_{\mathrm{b}}}
\newcommand{\ubasin}{u_{\mathrm{b}}}

\newcommand{\hplat}{h_{\mathrm{p}}}
\newcommand{\vplat}{v_{\mathrm{p}}}

\newcommand{\gamint}{\gamma_{\mathrm{int}}}
\newcommand{\gamext}{\gamma_{\mathrm{ext}}}
\newcommand{\gaminthat}{\hat{\gamma}_{\mathrm{int}}}
\newcommand{\gamexthat}{\hat{\gamma}_{\mathrm{ext}}}

\begin{document}

\title{Fluctuations of non-ergodic stochastic processes}

\author{G. George
\and L. Klochko
\and A.N. Semenov
\and J. Baschnagel
\and J.P.~Wittmer\thanks{joachim.wittmer@ics-cnrs.unistra.fr}
}
\institute{Institut Charles Sadron, Universit\'e de Strasbourg \& CNRS, 23 rue du Loess, 67034 Strasbourg Cedex, France}
\date{Received: date / Revised version: date}

\abstract{We investigate the standard deviation $\delta v(\tsamp)$ of the variance $v[\xbf]$ of time series $\xbf$ 
measured over a finite sampling time $\tsamp$ focusing on  non-ergodic systems 
where independent ``configurations" $c$ get trapped in meta-basins of a generalized phase space.
It is thus relevant in which order averages over the configurations $c$
and over time series $k$ of a configuration $c$ are performed.
Three variances of $v[\xbf_{ck}]$ must be distinguished: 
the total variance $\dvtot = \dvint + \dvext$
and its contributions $\dvint$, the typical internal variance within the meta-basins,
and $\dvext$, characterizing the dispersion between the different basins.
We discuss simplifications for physical systems where the stochastic variable $x(t)$ 
is due to a density field averaged over a large system volume $V$.
The relations are illustrated for the shear-stress fluctuations in quenched elastic networks
and low-temperature glasses formed by polydisperse particles and free-standing polymer films.
The different statistics of $\svint$ and $\svext$ are manifested by their different system-size dependences.}
\maketitle

\section{Introduction}
\label{sec_intro}

\begin{figure}[t]
\centerline{\resizebox{.65\columnwidth}{!}{\includegraphics*{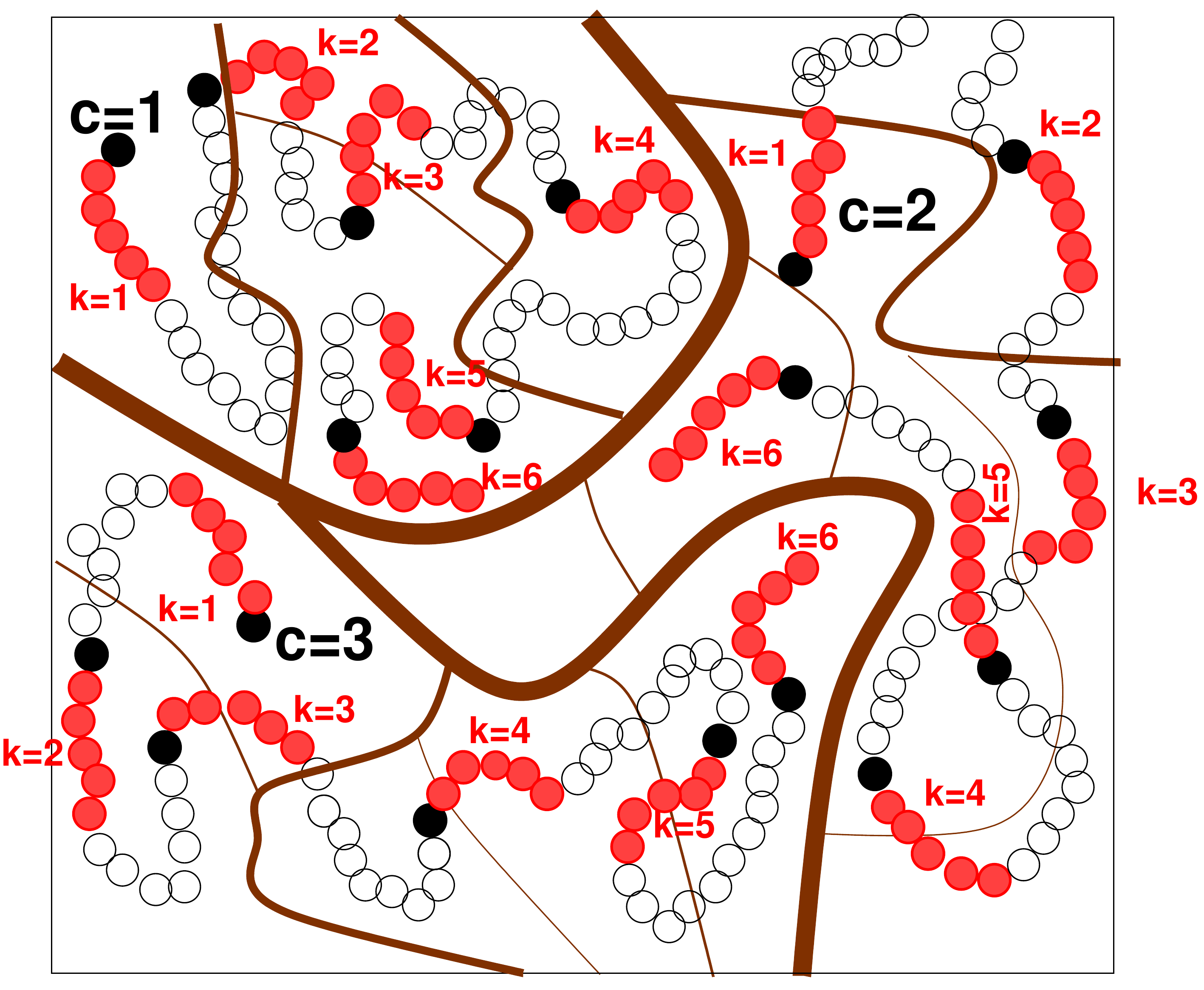}}}
\caption{Sketch of problem:
Time series $\xbf$ with $\Nt=6$ data entries $x_i$ are marked by filled circles.
The first entry $x_{i=1}$ is indicated by a dark filled circle.
The open circles mark tempering steps between different time series $k$ of
each {\em independently} prepared ``configuration" $c$.
The solid lines mark barriers of different height in some phase space.
We assume that the system is non-ergodic, i.e. 
the configurations $c$ are permanently trapped in the meta-basins 
marked by the thickest lines.
\label{fig_intro_sketch}
}
\end{figure}

Expectation values $\Ocal$ and standard deviations $\delta \Ocal$ of properties $\Ocal[\xbf]$ 
averaged over finite time series $\xbf$ of stochastic processes $x(t)$ 
\cite{numrec,vanKampenBook} are of relevance for a large variety of problems in 
scientific computing in general \cite{numrec,TaoPang} and especially in condensed matter 
\cite{ChaikinBook,DoiEdwardsBook,RubinsteinBook,HansenBook,FerryBook,GraessleyBook},
material modeling \cite{TadmorCMTBook,TadmorMMBook} 
and computational physics \cite{AllenTildesleyBook,LandauBinderBook}.
We consider ensembles of equidistant time series 
\begin{equation}
\xbf = \{x_i=x(t_i = i \tincr),i=1,\ldots,\Nt\}
\label{eq_xbf_def}
\end{equation}
each containing $\Nt$ data entries $x_i$.  
The data sequence is taken from $t_1=\tincr$ up to the ``sampling time" $\tsamp = \Nt \tincr$.\footnote{The 
term ``sampling time" is elsewhere often used for the time-interval $\tincr$ between neighboring data points.}
Examples of such time series obtained in a generalized phase space are sketched in Fig.~\ref{fig_intro_sketch}.
If the stochastic process $x(t)$ is {\em stationary} it may be characterized by means of 
the mean-square displacement
\begin{equation}
h(|t_i-t_j|) \equiv h_{i-j} \equiv \la (x_i-x_j)^2 \ra/2,
\label{eq_ht_def}
\end{equation}
of the data entries $x_i$. Note that $h(t) = c(0)-c(t)$ is closely related to the common 
autocorrelation function (ACF) $c(t)= \la x(t) x(0) \ra$ \cite{HansenBook}.\footnote{The 
response function due to an externally applied ``force" conjugated to $x$ 
switched on at $t=0$ is given within linear response by $h(t)$ \cite{DoiEdwardsBook}.}
Ensemble averages $\la \ldots \ra$ are commonly estimated by ``{\em $c$-averaging}" over 
many independently prepared systems $c$, called here ``configurations". 
An example with $\Nc=3$ is given in Fig.~\ref{fig_intro_sketch}.
As in our previous work \cite{lyuda19a,spmP1},
we shall focus on the ``empirical sample variance"\footnote{$v[\xbf]$ 
is defined without the usual ``Bessel correction" \cite{numrec}.
See the discussion at the end of Sec.~\ref{theo_definitions}.}
\begin{equation}
\Ocal[\xbf] = v[\xbf] \equiv 
\frac{1}{\Nt} \sum_{i=1}^{\Nt} (x_i -\overline{x})^p \mbox{ with } \overline{x} = \frac{1}{\Nt} \sum_{i=1}^{\Nt} x_i
\label{eq_vxdef}
\end{equation}
and $p=2$.
Importantly, its expectation value $v = \la v[\xbf] \ra$ and 
variance $\delta v^2 = \la (v[\xbf]-v)^2 \ra$ are given by \cite{lyuda19a,spmP1} 
\begin{eqnarray}
v         & = & \frac{2}{\Nt^2} \sum_{i=1}^{\Nt-1} (\Nt-i) \ h_i \mbox{ and } \label{eq_bg_1} \\ 
\delta v^2 & = & \dvgauss[h] \equiv \frac{1}{2\Nt^4} \sum_{i,j,k,l=1}^{\Nt} \ g_{ijkl}^2 \ \mbox{ with } \nonumber \\
g_{ijkl} & \equiv & (h_{i-j} + h_{k-l}) - (h_{i-l} + h_{j-k})
\label{eq_bg_2}
\end{eqnarray}
in terms of the ACF $h(t)$. While Eq.~(\ref{eq_bg_1}) is a direct consequence
of the {\em stationarity} of the process, Eq.~(\ref{eq_bg_2}) assumes in addition that $x(t)$ is
both {\em Gaussian} and {\em ergodic} \cite{spmP1}. 
Note that $v$ and $\delta v$ depend in general on the sampling time $\tsamp$ of the time 
series.\footnote{As seen by analyzing Eq.~(\ref{eq_bg_2}) \cite{lyuda19a,spmP1},
the standard deviation $\delta v(\tsamp)$ is small
if $h(t)$ is essentially constant for $t \approx \tsamp$ but may become of order of $v(\tsamp)$
if $h(t)$ changes strongly for $t \approx \tsamp$.}

As sketched by the thickest solid lines in Fig.~\ref{fig_intro_sketch}, 
if some large barriers are present in the generalized phase space the stochastic processes of 
independent configurations $c$ must get trapped in meta-basins \cite{Heuer08,Gardner}, 
at least for sampling times $\tsamp \ll \taualph$ with $\taualph$
being the terminal relaxation time of the system.
For such non-ergodic systems and for sufficiently large sampling times $\tsamp$ (to be specified below)
it was found \cite{Procaccia16,lyuda19a,spmP1} that 
$\delta v(\tsamp)$ becomes similar to a constant ``non-ergodicity parameter" $\Snonerg > 0$. 
$\delta v$ thus differs from the rapidly decaying Gaussian prediction 
$\svgauss \propto 1/\sqrt{\tsamp}$ \cite{lyuda19a,spmP1}. 
To understand the observed discrepancy an {\em extended ensemble} of time series $\xbf_{ck}$ is 
needed where for each configuration $c$ one samples $\Nk\gg 1$ time series $k$.\footnote{The 
time series $k$ may be obtained by first tempering
the configuration $c$ over a time interval $\ttemper$ larger than the
basin relaxation time $\taubasin$
and by sampling then $\Nk$ time intervals $\tsamp$ separated
by constant spacer intervals $\tspacer \gg \taubasin$.}
$k$-averages and $k$-variances may then depend on the configuration $c$
and it becomes relevant in which order $c$-averages over configurations $c$
and $k$-averages over time series $k$ of a given configuration $c$ are performed.
As described in Sec.~\ref{theo_ck}, three variances must be distinguished: 
\begin{itemize}
\item
the standard ``total variance" $\dvtot(\tsamp)$ obtained 
by lumping together the quantities $v[\xbf_{ck}]$ for all $c$ and $k$,
\item
the $c$-averaged ``internal variance" $\dvint(\tsamp,\Nk)$ of the meta-basins and
\item
the ``external variance" $\dvext(\tsamp,\Nk)$ describing the dispersion between the different meta-basins.
\end{itemize}
$\svtot$ is commonly probed in previous computational work on fluctuations of $v$ 
\cite{Procaccia16,WKC16,ivan17c,ivan18,film18,lyuda19a,spmP1}.
Importantly,
\begin{equation}
\dvtot(\tsamp) = \dvint(\tsamp,\Nk) + \dvext(\tsamp,\Nk)
\label{eq_key_1}
\end{equation}
holds rigorously for large $\Nc$ and the $\Nk$-dependence 
on the right-hand side becomes rapidly irrelevant with increasing $\Nk$ for non-ergodic systems.
As will be discussed in Sec.~\ref{theo_dv}, Eq.~(\ref{eq_key_1}) can be simplified in many cases 
such that the total variance $\svtot(\tsamp)$ can be traced back to the ACF $h(t)$
and the ``non-ergodicity parameter" $\Snonerg$ properly defined in Sec.~\ref{theo_ck}. 
This leads especially to
\begin{equation}
\svint(\tsamp) \simeq \sqrt{\taubasin/\tsamp} \mbox{ and } \svext(\tsamp) \simeq \Snonerg
\label{eq_key_2}
\end{equation}
for $\tsamp \gg \taubasin$ with $\taubasin$ being the typical basin relaxation time.
Corroborating Ref.~\cite{spmP1} it will be seen that system-size effects become rapidly irrelevant
for physical systems where $x(t)$ is the average over a statistically uniform density field (Sec.~\ref{theo_V}).

Various relations and issues discussed theoretically in  Sec.~\ref{sec_theo}
are tested numerically in Sec.~\ref{sec_shear} for the fluctuations of the shear stresses in three 
strictly or in practice non-ergodic coarse-grained model systems described in Sec.~\ref{sec_algo}.
Temperature-effects are briefly discussed in Sec.~\ref{shear_T},
sys\-tem-size effects in Sec.~\ref{shear_V}.
The paper concludes in Sec.~\ref{sec_conc} with a summary and an outlook to future work.
Appen\-dix~\ref{app_V} presents further details on the power-law exponents describing
the system-size dependence of $v$ and $\Snonerg$,
Appendix~\ref{app_px} the distribution of the frozen $v_c$ for different configurations $c$.

\section{Theoretical considerations}
\label{sec_theo}

\subsection{Some notations}
\label{theo_definitions}

To state compactly the expressions developed below it is useful to introduce a few notations. 
The $l$-average operator 
\begin{equation}
\Eop^l \Ocal_{lmn\ldots}  \equiv \frac{1}{\Nl} \sum_{l=1}^{\Nl} \Ocal_{lmn\ldots} \equiv \Ocal_{mn\ldots}(\Nl) \label{eq_Eopdef}
\end{equation}
takes a property $\Ocal_{lmn\ldots}$ depending possibly on several indices $l,m,\ldots$
and projects out the specified index $l$, 
i.e. the generated property $\Ocal_{mn\ldots}(\Nl)$ does not depend any more on $l$,
but it may depend on the upper bound $\Nl$ as marked by the argument.
The latter dependence drops out for large $\Nl$ (formally $\Nl \to \infty$) 
if $\Ocal_{lmn\ldots}$ is stationary or converges with respect to $l$.
The $l$-variance operator $\Vop^l$ is {\em defined} by
\begin{equation}
\Vop^l \Ocal_{lmn\ldots} \equiv \frac{1}{\Nl} \sum_{l=1}^{\Nl} \left( \Ocal_{lmn\ldots}-\Eop^l \Ocal_{lmn\ldots} \right)^2.
\label{eq_Vopdef} 
\end{equation}
Introducing the power-law operator $\Pop^{\alpha} \Ocal \equiv \Ocal^{\alpha}$,
with the exponent $\alpha=2$ being here the only relevant case,
and using the standard commutator $[\Aop,\Bop] \equiv \Aop \Bop - \Bop \Aop$ 
for two operators $\Aop$ and $\Bop$,
the $l$-variance operator may be written $\Vop^l = [\Eop^l,\Pop^2]$.
The result $\delta \Ocal_{mn\ldots}^2(\Nl) = \Vop^l \Ocal_{lmn\ldots}$ of this operation
on $\Ocal_{lmn\ldots}$ depends in general on the upper bound $\Nl$.
In the cases considered below $\delta \Ocal_{mn\ldots}^2(\Nl)$ converges for large $\Nl$
and the $\Nl$-dependency again drops out. This large-$\Nl$ limit is written 
\begin{equation}
\delta \Ocal_{mn\ldots}^2(\ldots) \equiv \lim_{\Nl\to \infty} \delta \Ocal_{mn\ldots}^2(\Nl,\ldots) 
\label{eq_Vl_large_Nl}
\end{equation}
where the dots $\ldots$ indicate possible additional variables.
We emphasize finally that we have defined the $l$-variance operator $\Vop^l$,
as above in Eq.~(\ref{eq_vxdef}) for $v[\xbf]$, 
as an ``uncorrected biased sample variance" without the often used Bessel correction \cite{numrec,LandauBinderBook}, 
i.e. we normalize with $1/\Nl$ and not with $1/(\Nl-1)$.
If the $\Nl$ contributions $l$ are {\em uncorrelated} this can be readily shown to underestimate 
the asymptotic variance by a factor of $(\Nl-1)/\Nl$ \cite{LandauBinderBook}, i.e.
\begin{equation}
\delta \Ocal_{mn\ldots}^2(\Nl,\ldots) = \left(1-\frac{1}{\Nl}\right) \delta \Ocal_{mn\ldots}^2(\ldots).
\label{eq_Vl_independent}
\end{equation}

\subsection{Extended ensembles of time series $\xbf_{ck}$}
\label{theo_ck}

\subsubsection{Ergodic systems}
\label{theo_ck_ergodic}

We remind first that in ergodic systems the terminal relaxation time $\taualph$ 
is short relative to reasonable experimental or computational sampling times $\tsamp$, 
i.e. the time series can easily cross all barriers.
One may thus either compute the averages $\Eop^c \Ocal[\xbf_c]$ and $\Vop^c \Ocal[\xbf_c]$ 
over $\Nc$ independent configurations $c$ (with $\Nk=1$)
or the averages $\Eop^k \Ocal[\xbf_k]$ and $\Vop^k \Ocal[\xbf_k]$ over $\Nk \gg 1$
different time series $k$ of one long trajectory (with $\Nc=1$). Hence, 
\begin{equation}
\Eop^c \Ocal[\xbf_c] \simeq \Eop^k \Ocal[\xbf_k] \mbox{ and } 
\Vop^c \Ocal[\xbf_c] \simeq \Vop^k \Ocal[\xbf_k]
\label{eq_ergodic}
\end{equation}
holds for sufficiently large $\Nc$ and $\Nk$.
Importantly, it is sufficient for ergodic systems to characterize a time series $\xbf$
by {\em one} index. We come back to ergodic systems in Sec.~\ref{theo_ck_ergodic2}.

\subsubsection{Non-Ergodic systems}
\label{theo_ck_nonergodic}

Let us focus now on strictly non-ergodic systems with infinite terminal relaxation times $\taualph$ 
for the transitions between the meta-basins.
We characterize a time series $\xbf_{ck}$ by the {\em two} discrete indices $c$ and $k$
with $1 \le c \le \Nc$ and $1 \le k \le \Nk$.
As shown in Fig.~\ref{fig_intro_sketch}, the index $c$ stands for the ``configurations" 
(or set-ups) generated by completely independent preparation histories for the system probed,
the index $k$ for subsets of length $\Nt$ of a much larger trajectory
generated for a fixed configuration $c$. 
The central point is now that 
\begin{eqnarray}
\Ocal_c(\tsamp,\Nk) & \equiv & \Eop^k \Ocal[\xbf_{ck}]
\mbox{ and } \label{eq_Ocal_c_def} \\
\delta \Ocal^2_c(\tsamp,\Nk)  & \equiv & \Vop^k \Ocal[\xbf_{ck}]
\label{eq_dvcal_c_def}
\end{eqnarray}
do depend in general not only on the sampling time $\tsamp = \Nt \tincr$
of the time series and the number $\Nk$ of time series probed but crucially also on $c$
--- even for arbitrarily large $\Nt$ and $\Nk$ --- 
since the ``$c$-trajectory" of each configuration $c$ is trapped
(Fig.~\ref{fig_intro_sketch}).
For $\tsamp \gg \taubasin$ much larger than the typical basin relaxation time $\taubasin$ 
the $\tsamp$-dependence of $\Ocal_c(\tsamp,\Nk)$ 
drops out and $\delta \Ocal_c(\tsamp,\Nk) \propto 1/\sqrt{\tsamp/\taubasin}$
since we average over $\tsamp/\taubasin$ independent subintervals.
Moreover, the $\Nk$-dependence must disappear if $\Nk \gg 1$ and
the $c$-trajectory has completely explored the basin.
Assuming that after each measurement interval of length $\tsamp$ a spacer (tempering) step of length $\tspacer$ follows,
as marked by the open circles in Fig.~\ref{fig_intro_sketch},
this happens for $c$-trajectories of total length $\tsampmax \equiv  \Nk \times (\tsamp+\tspacer)$ with 
\begin{equation}
\taubasin \ll \tsamp \ll \tsampmax \ll \taualpha.
\label{eq_tsampmax}
\end{equation} 
The first inequality implies that the sampling is ergodic within the metabasin 
(that's why, the metabasin is sometimes said to be an "ergodic component"), 
while the last inequality states the ergodicity breaking of the system.

\subsubsection{Commuting and non-commuting operators}
\label{theo_ck_commute}

Since $[\Eop^c,\Eop^k]=0$ we may write quite generally 
\begin{equation}
\Eop^c \Eop^k \ \Ocal[\xbf_{ck}] = \Eop^k \Eop^c \ \Ocal[\xbf_{ck}] = 
\Eop^l \ \Ocal[\xbf_l] = \Ocal,
\label{eq_etc}
\end{equation}
i.e. the two indices $c$ and $k$ can be lumped together to one index $l$.
Averages of this type are called ``simple averages".
For instance, the average variance $v = \Eop^c \Eop^k v[\xbf_{ck}] = \Eop^l v[\xbf_l]$ 
is a simple average. 
At variance to this in general
\begin{equation}
[\Eop^c,\Vop^k] \ne 0 \mbox{ or } [\Vop^c,\Eop^k] \ne 0 \mbox{ if } \Nk > 1.
\label{eq_non_commute}
\end{equation}
Two operators of this type thus cannot be commuted and the indices $c$ and $k$
cannot be exchanged or lumped together.

\subsubsection{Different variances}
\label{theo_ck_variances}

We define now in general terms the three variances mentioned in the Introduction:
\begin{eqnarray}
\dOtot(\tsamp,\Nc,\Nk) & \equiv & [\Eop^c \Eop^k,\Pop^2] \Ocal[\xbf_{ck}] \label{eq_dOtot} \\
\dOint(\tsamp,\Nc,\Nk) & \equiv & \Eop^c \delta \Ocal_c^2 = \Eop^c \Vop^k \Ocal[\xbf_{ck}] \label{eq_dOint} \\
\dOext(\tsamp,\Nc,\Nk) & \equiv & \Vop^c \Ocal_c = \Vop^c \Eop^k \Ocal[\xbf_{ck}].    \label{eq_dOext}
\end{eqnarray}
The indicated dependencies on $\tsamp$, $\Nc$ and $\Nk$ will be discussed in detail below
(Sec.~\ref{theo_ck_Nc}-\ref{theo_ck_large_tsamp}).
Let us stress first that the ``total variance" $\dOtot$ is a simple average, i.e. all time series $\xbf_{ck}$ 
can be lumped together:
\begin{equation}
\dOtot = \Vop^l \Ocal[\xbf_l] = [\Eop^l,\Pop^2] \Ocal[\xbf_l].
\label{eq_dOtot_lumped}
\end{equation}
Importantly, the expectation value of $\sOtot$ for $\Nc \to \infty$ is strictly $\Nk$-independent and 
may be also computed by using only {\em one} time series for each configuration ($\Nk=1$).
$\dOtot$ is thus the standard commonly computed variance 
\cite{Procaccia16,WKC16,ivan17c,ivan18,film18,lyuda19a,spmP1}.
The ``internal variance" $\dOint$ and the ``external variance" $\dOext$ are different types of observables 
since Eq.~(\ref{eq_non_commute}) holds, i.e. $c$ and $k$ cannot be lumped together.
Note also that $\sOint$ and $\sOext$ do depend on $\Nk$ even for $\Nc \to \infty$
and that $\sOext$ vanishes if all $\Ocal_c$ are identical.
Using the identity
\begin{eqnarray}
\Vop^l & = & [\Eop^l,\Pop^2] =  [\Eop^c \Eop^k,\Pop^2] \nonumber \\
& = &\Eop^c \Eop^k \Pop^2 - \Eop^c \Pop^2 \Eop^k + \Eop^c \Pop^2 \Eop^k - \Pop^2 \Eop^c \Eop^k \nonumber \\
& = & \Eop^c \Vop^k + \Vop^c \Eop^k
\label{eq_opschain}
\end{eqnarray}
$\dOtot$ can be {\em exactly} decomposed as the sum 
\begin{eqnarray}
\dOtot(\tsamp,\Nc,\Nk) & = & \dOint(\tsamp,\Nc,\Nk) \nonumber \\
                   & + & \dOext(\tsamp,\Nc,\Nk)
\label{eq_key_1_gen}
\end{eqnarray}
of the two {\em independent} variances $\dOint$ and $\dOext$.
Details of both contributions $\sOint$ and $\sOext$ depend on the properties of the 
considered stochastic process $x(t)$ and the functional $\Ocal[\xbf]$ considered.
However, the following fairly general statements can be made.

\subsubsection{$\Nc$-dependences}
\label{theo_ck_Nc}

Let us define the large-$\Nc$ limits
\begin{eqnarray}
\sOext(\tsamp,\Nk) & \equiv & \lim_{\Nc\to\infty} \sOext(\tsamp,\Nk,\Nc) \label{sOext_largeNc}\\
\sOtot(\tsamp) & = & \sOtot(\tsamp,\Nk) \nonumber \\
& \equiv & \lim_{\Nc\to\infty} \sOtot(\tsamp,\Nk,\Nc) \label{sOtot_largeNc}
\end{eqnarray}
where the $\Nk$-dependence of $\sOtot$ does not emerge as already stated below Eq.~(\ref{eq_dOtot_lumped}).
As all the configurations $c$ are assumed to be strictly independent, 
$\sOint$ does not depend on $\Nc$, i.e. 
\begin{eqnarray}
\sOint(\tsamp,\Nc,\Nk) & = & \sOint(\tsamp,\Nk) \mbox{, and } \label{eq_dOintA} \\
\dOext(\tsamp,\Nc,\Nk) & = & \left(1-\frac{1}{\Nc}\right) \dOext(\tsamp,\Nk) \label{eq_dOextA}
\end{eqnarray}
where we have used the general relation Eq.~(\ref{eq_Vl_independent}). 
Using Eq.~(\ref{eq_key_1_gen}) this implies 
\begin{equation}
\dOtot(\tsamp,\Nc,\Nk) = \dOtot(\tsamp) - \frac{\dOext(\tsamp,\Nk)}{\Nc}.
\label{eq_key_1_genB}
\end{equation}
If not emphasized otherwise, we assume below that $\Nc$ is large, say at least $\Nc \approx 100$,
and the stated $\Nc$-dependences thus become irrelevant.

\subsubsection{$\Nk$-dependences}
\label{theo_ck_Nk}

While $\sOint$ and $\sOext$ depend in principle on $\Nk$, this dependence must drop out 
for large $\Nk$ if $\tsampmax \gg \taubasin$ as noted in Sec.~\ref{theo_ck_variances}.
It is therefore useful to define:
\begin{eqnarray}
\sOint(\tsamp) & \equiv & \lim_{\Nk \to \infty} \sOint(\tsamp,\Nk), \label{eq_sOint_largeNk} \\
\sOext(\tsamp) & \equiv & \lim_{\Nk \to \infty} \sOext(\tsamp,\Nk). \label{eq_sOext_largeNk} 
\end{eqnarray}
Note also that $\sOint(\tsamp,\Nk)=0$ and $\sOext(\tsamp,\Nk) = \sOtot(\tsamp)$ 
in the opposite limit, $\Nk = 1$.
In what follows we assume that the spacer time intervals $\tspacer$ 
between the measured time series $k$ of a configuration $c$ is large,
i.e. either $\tspacer \gg \taubasin$ or $\tspacer + \tsamp \gg \taubasin$.
In this case all $\Nk$ time series for each configuration must be virtually independent
(albeit constraint to be in the same basin).
Therefore, 
\begin{equation}
\dOint(\tsamp,\Nk) \simeq \left( 1-\frac{1}{\Nk} \right) \ \dOint(\tsamp)
\label{eq_dOint_Nk}
\end{equation}
providing the $\Nk$-dependence of $\sOint$ for sufficiently large $\tspacer$. 
Using Eq.~(\ref{eq_key_1_gen})
both for finite $\Nk$ and for $\Nk \to \infty$ and the fact that 
$\sOtot(\tsamp,\Nk)=\sOtot(\tsamp)$, i.e. $\sOtot$ does not depend on $\Nk$ for large $\Nc$, 
we get 
\begin{equation}
\dOext(\tsamp,\Nk) \simeq \dOext(\tsamp) + \frac{1}{\Nk} \ \dOint(\tsamp)
\label{eq_dOext_Nk}
\end{equation}
for $\tspacer \gg \taubasin$ and $\Nc \to \infty$.
$\sOext(\tsamp,\Nk)$ thus depends on $\Nk$ and $\sOext(\tsamp)$ and,
interestingly, also on $\sOint(\tsamp)$.
 
\subsubsection{Total variance $\dvtot(\tsamp,\Nc,\Nk)$}
\label{theo_ck_svtot}

Using Eqs.~(\ref{eq_key_1_genB}, \ref{eq_dOext_Nk}) 
the total variance, Eq.~(\ref{eq_key_1_gen}), can be written for finite $\Nc$ as
\begin{eqnarray}
&&\dOtot(\tsamp,\Nc,\Nk) \simeq \left(1-\frac{1}{\Nk}\right) \dOint(\tsamp) \nonumber \\
& + & \left(1-\frac{1}{\Nc}\right) \left(\dOext(\tsamp) + \frac{1}{\Nk} \dOint(\tsamp) \right).
\label{eq_dOtot_Nk}
\end{eqnarray}
The latter equation is valid for $\tsamp + \tspacer \gg \taubasin$ and $\tsampmax \ll \taualph$.
It shows explicitly how $\dOtot$ depends on the number of configurations $\Nc$ and the
number of time series $\Nk$ for each $c$.
For $\Nc \to \infty$ Eq.~(\ref{eq_dOtot_Nk}) simplifies to 
\begin{eqnarray}
\dOtot(\tsamp,\Nc,\Nk) &\to & \dOtot(\tsamp) \nonumber \\
&= & \dOint(\tsamp) + \dOext(\tsamp)
\label{eq_dOtot_Nk_largeNc}
\end{eqnarray}
i.e. as expected from Sec.~\ref{theo_ck_variances}
not only the $\Nc$-depen\-dence but also the $\Nk$-dependence drops out.

\subsubsection{Large-$\tsamp$ limit $(\tsamp \gg \taubasin)$}
\label{theo_ck_large_tsamp}

Here and below we return to real non-ergodic systems
with very large but finite terminal relaxation times $\taualph$.
Without additional assumptions it is also clear that 
\begin{equation}
\sOint \propto 1/\sqrt{\tsamp/\taubasin}, \ \sOext(\tsamp) \simeq \Snonerg = const,
\label{eq_large_tsamp}
\end{equation}
for $\taualph \gg \tsamp \gg \taubasin$
with the ``non-ergodicity parameter" $\Snonerg$ being defined by the 
finite limit of $\sOext$ at large $\tsamp$
\begin{equation}
\Snonerg \equiv \lim_{\tsamp/\taubasin \to \infty} \sOext(\tsamp,\Nk).
\label{eq_Snonerg_def}
\end{equation}
This is equivalent to the large-$\tsamp$ limit of $\sOtot(\tsamp)$
since the $\Nk$-dependence of $\sOext$ drops out for large $\tsamp$.
(The last statement may be also seen from Eq.~(\ref{eq_dOext_Nk}).)
As already noted, the first asymptotic law in Eq.~(\ref{eq_large_tsamp}) is a consequence 
of the $\tsamp/\taubasin$ uncorrelated subintervals
for each $c$-trajectory while the second limit is merely a consequence of the $\Ocal_c(\tsamp)$ becoming constant.
Equation~(\ref{eq_large_tsamp}) implies that $\sOtot$ must become
\begin{equation}
\sOtot \to \sOext \approx \Snonerg \mbox{ for } \tsamp \gg \Tnonerg \gg \taubasin.
\label{eq_sOtot_large_tsamp}
\end{equation}
Note that the crossover to the $\Snonerg$-dominated regime occurs at an additional time scale $\Tnonerg$. 
Operationally, this ``non-ergodicity time" $\Tnonerg$ may be defined as
\begin{equation}
\sOint(\tsamp \stackrel{!}{=}\Tnonerg) = \Snonerg.
\label{eq_Tnonerg_def}
\end{equation}
$\Snonerg$ does not dependent on $\Nk$, 
being equivalently the large-$\tsamp$ limit of either $\sOext(\tsamp,\Nk)$ or $\sOtot(\tsamp)$, 
the latter simple average being strictly $\Nk$-independent ($\Nc \to \infty$).
Coming back to Eq.~(\ref{eq_dOext_Nk}) and using Eq.~(\ref{eq_Snonerg_def})
one sees that
\begin{eqnarray}
\dOext(\tsamp,\Nk) & \simeq & \Dnonerg + \frac{1}{\Nk} \dOint(\tsamp) \ \mbox{and}  
\label{eq_dOext_Nk_large_tsamp} \\
\dOtot(\tsamp) & \simeq & \Dnonerg + \dOint(\tsamp) \label{eq_dOtot_large_tsamp}
\end{eqnarray}
for $\taualph \gg \tsamp \gg \taubasin$ and $\Nc \to \infty$.

\subsubsection{Back to ergodic systems}
\label{theo_ck_ergodic2}

Let us finally assume that the terminal relaxation time $\taualph$ is 
shorter than the sampling time, $\tsamp \gg \taualph$.
In this ergodic limit all trajectories become statistically equivalent,
i.e. $\sOext(\tsamp) = 0$ (cf. Eq.~(\ref{eq_sOext_largeNk})).
Following Eq.~(\ref{eq_dOextA}) and Eq.~(\ref{eq_dOext_Nk}) we have
\begin{equation}
\dOext(\tsamp,\Nc,\Nk) = \left(1-\frac{1}{\Nc}\right) \frac{1}{\Nk} \dOint(\tsamp)
\label{eq_dOext_ergodic}
\end{equation}
and using Eq.~(\ref{eq_dOintA}) and Eq.~(\ref{eq_dOint_Nk}) we get 
\begin{equation}
\dOint(\tsamp,\Nc,\Nk) = \left(1-\frac{1}{\Nk}\right) \dOint(\tsamp).
\label{eq_dOint_ergodic}
\end{equation}
This implies by means of Eq.~(\ref{eq_key_1_gen}) or, equivalently, using Eq.~(\ref{eq_dOtot_Nk})
\begin{equation}
\dOtot(\tsamp,\Nc,\Nk) = \left(1-\frac{1}{\Nk\Nc}\right) \dOint(\tsamp).
\label{eq_dOtot_ergodic}
\end{equation}
For either $\Nc \to \infty$ or $\Nk \to \infty$
the latter relation yields finally 
\begin{equation}
\sOtot(\tsamp,\Nc,\Nk) \to \sOtot(\tsamp)=\sOint(\tsamp)
\label{eq_dOtot_ergodic_largeNc}
\end{equation}
which is similar to the second relation stated in Eq.~(\ref{eq_ergodic}).

\subsection{Properties related to $\Ocal[\xbf]=v[\xbf]$}
\label{theo_dv}

From now on we shall focus on $\Ocal[\xbf]=v[\xbf]$, Eq.~(\ref{eq_vxdef}), for $p=2$.
Our key results Eq.~(\ref{eq_key_1}) and Eq.~(\ref{eq_key_2}) follow directly from 
the more general relations Eq.~(\ref{eq_key_1_gen}) and Eq.~(\ref{eq_large_tsamp}).
Assuming an {\em ergodic} Gaussian process we have expressed $\delta v(\tsamp)$
by the functional $\svgauss[h]$ in terms of the ACF $h$, Eq.~(\ref{eq_bg_2}).
Numerically better behaved equivalent reformulations are discussed in Ref.~\cite{spmP1}.
We make now the additional physical assumption that 
\begin{quote}
{\em after sufficient tempering the stochastic process of each configuration $c$ in its meta-basin
is both stationary and Gaussian}. 
\end{quote}
This implies that for $\taubasin \ll \tsampmax \ll \taualph$
Eq.~(\ref{eq_bg_2}) may hold for each basin separately.\footnote{This 
assumption also holds for $\tsamp \gg \taualph$ for a finite terminal relaxation time $\taualph$ 
associated with the transitions between the meta-basins. Note that the systems is ergodic in the second regime.}
i.e. $\delta v_c$ is given by $\svgauss[h_c]$ expressed in terms of the corresponding 
ACF $h_c$ of the basin instead of its $c$-average $h = \Eop^c h_c$.
Unfortunately, $h_c$ is not known in general (at least not to sufficient accuracy), 
but rather $h$. Since Eq.~(\ref{eq_bg_2}) corresponds to products of $h_c$, 
it is a ``mean-field type" approximation to replace $h_c$ by its $c$-average $h$.
This technical assumption becomes strictly valid for large systems, $V \to \infty$,
since fluctuations of the ACF vanish in this limit.
Within the above physical assumption and the additional technical approximation one thus expects after a final $c$-averaging 
\begin{equation}
\svint(\tsamp) \approx \svgauss[h] \mbox{ with } h = \Eop^c h_c
\label{eq_svgauss2svint}
\end{equation}
to hold for all $\tsamp$.
Whether this approximation is good enough must be checked for each case.
Note that neither $\svgauss[h]$ nor $\svint(\tsamp)$ do depend (explicitly) on $\Nc$ or $\Nk$,
i.e. Eq.~(\ref{eq_svgauss2svint}) only holds for $\svint(\tsamp,\Nk)$ with sufficiently large $\Nk$.
Fortunately, due to Eq.~(\ref{eq_dOint_Nk}) 
\begin{equation}
\svint(\tsamp) \simeq \svint(\tsamp,\Nk)/\sqrt{1-1/\Nk},
\label{eq_dv_svint}
\end{equation}
i.e. by computing even a small number $\Nk$ of time series the asymptotic limit $\svint(\tsamp)$ may be obtained.
The relations Eq.~(\ref{eq_dOtot_Nk_largeNc}), Eq.~(\ref{eq_Snonerg_def}) and
Eq.~(\ref{eq_svgauss2svint}) suggest the simple interpolation 
\begin{equation}
\svtot(\tsamp) \approx \sqrt{\dvgauss[h] + \Snonerg^2}
\label{eq_svtot_intpol}
\end{equation}
stating that $\svtot$ is essentially given by $h(t)$ plus an additional constant $\Snonerg$.

\subsection{General system-size effects}
\label{theo_V}

The stochastic processes of many systems are to a good approximation Gaussian 
since the data entries $x_i = \Eop^m x_{im}$ are averages over $\Nm \gg 1$ microscopic contributions $x_{im}$
and the central limit theorem applies \cite{vanKampenBook}. 
(These contributions are often unknown and experimentally inaccessible.)
It is assumed here that the system is split in $\Nm$ quasi-independent microcells,
$\Nm$ is proportional to the volume $V$, and $x_{im}$ comes from the $m$-th microcell.
Albeit the $x_{im}$ may be correlated, i.e. they may not all fluctuate independently,
the fluctuations of the $x_i$ commonly decrease with increasing $\Nm$.
As a consequence, $\svint$ and $\svext$ generally decrease with the system size.
Assuming scale-free correlations one may write \cite{spmP1}
\begin{equation}
\svint(\tsamp) \propto 1/\Nm^{\gaminthat} \mbox{ and } \svext(\tsamp) \propto 1/\Nm^{\gamexthat}
\label{eq_gamma_def}
\end{equation}
introducing the two phenomenological exponents $\gaminthat$ and $\gamexthat$.
If the stochastic processes of all basins are Gaussian the same exponent $\gaminthat$ must hold for 
$\svgauss[h]\approx \svint(\tsamp)$, Eq.~(\ref{eq_svgauss2svint}).
In turn due to Eq.~(\ref{eq_bg_2}) this implies the same exponent for $h(t)$ and 
then due to the stationarity relation Eq.~(\ref{eq_bg_1}) also for $v(\tsamp)$. 
Due to the definition Eq.~(\ref{eq_Snonerg_def}) 
the same exponent $\gamexthat$ must hold for $\svext(\tsamp)$ and $\Snonerg$.
  
As reminded in Appendix~\ref{app_V} it is readily seen that $\gaminthat = 1$ and $\gamexthat = 3/2$
for strictly uncorrelated variables $x_{im}$.
The uncorrelated reference with $\gaminthat=1$ is often included into the definition of the data entries 
by rescaling $x_i \Rightarrow \sqrt{\Nm} x_i$. 
Hence, $\gaminthat \Rightarrow \gamint \equiv \gaminthat-1$ and $\gamexthat \Rightarrow \gamext \equiv \gamexthat-1$ in the above relations, i.e. 
\begin{equation}
\gamint = 0 \mbox{ and } \gamext = 1/2 \label{eq_gamma_uncorr}
\end{equation}
for rescaled uncorrelated variables $x_{im}$.
Using the definition of the non-ergodicity time $\Tnonerg$, Eq.~(\ref{eq_Tnonerg_def}), 
and the asymptotic limit Eq.~(\ref{eq_large_tsamp}) it is seen that
\begin{equation}
\Tnonerg \propto \Nm^{2(\gamext-\gamint)} \mbox{ for } \Tnonerg \gg \taubasin. \label{eq_Tnonerg_V}
\end{equation}
For uncorrelated microcells we have $\Tnonerg \propto \Nm$ and, moreover,
$h(t)$ and thus $\taubasin$ are $\Nm$-independent, i.e. the condition $\Tnonerg \gg \taubasin$
becomes rapidly valid. 

\subsection{Fields of intensive thermodynamic variables}
\label{theo_T}

Up to now our description of ergodic and non-ergodic stochastic processes
has remained deliberately general and we have specifically avoided the notions 
and assumptions of thermodynamics and statistical physics \cite{ChaikinBook,TadmorCMTBook,TadmorMMBook}.
We shall now assume that each $c$-trajectory in its meta-basin 
is not only stationary and Gaussian but, moreover, at thermal equilibrium
albeit under the (not necessarily known) constraints imposed to the basin.

We focus below on (instantaneous) intensive thermodynamic variables $\tauhat$ 
(other than the temperature) which are $d$-dimensional volume averages
\begin{equation}
\tauhat(t) = \frac{1}{V} \int \ddiff \rvec \ \tauhat_{\rvec}(t)
\label{eq_volaver}
\end{equation}
over (instantaneous) fields $\tauhat_{\rvec}(t)$ of local contributions (of same dimension).
For such generic fields $\Nm$ corresponds to the number of local volume elements $dV$ computed.
Following the rescaling convention mentioned in Sec.~\ref{theo_V} 
the stochastic process is obtained by rescaling
\begin{equation}
\tauhat(t) \Rightarrow x(t) \equiv \sqrt{\beta V} \tauhat(t)
\label{eq_rescale}
\end{equation}
with $\beta=1/\kB T$ being the inverse temperature.
For density fields $\tauhat_{\rvec}$ characterized by a finite correlation length $\xi$
this rescaling leads to the same exponents $\gamint=0$ and $\gamext=1/2$
as for completely uncorrelated microscopic variables.
This assumes that $\xi^d \ll V$ and that $\xi$ is $V$-independent.

Importantly, $\gamint=0$ must even hold for systems with some long-range correlations 
if standard thermostatistics can be used for each basin.
To see this let us first note that the large-$\tsamp$ limit $v_c$ of $v_c(\tsamp)$
is equivalent to the thermodynamically averaged variance of $x(t)$ for the basin.\footnote{The
stochastic process is ergodic within the basin.} Using the standard 
relation for the fluctuation of intensive thermodynamic variables \cite{ChaikinBook,WXP13} 
this implies that $v_c$ does not depend explicitly on $V$.\footnote{Albeit
$v_c$ depends on whether the average intensive variable $\sigma$ of the meta-basin is imposed 
or its conjugated extensive variable in both cases $v_c$ does not depend on $V$.
See Ref.~\cite{Lebowitz67} or Sec.~II.A of Ref.~\cite{WXP13} for details.} 
This suggests that $\gamint=0$ not only holds for $v_c$ but also for $v_c(\tsamp)$ and $v(\tsamp)=\Eop^c v_c(\tsamp)$
and in turn 
using Eq.~(\ref{eq_bg_1}) also for $h_c(t)$ and $h(t)=\Eop^ch_c(t)$,
using Eq.~(\ref{eq_bg_2}) also for $\svgauss[h_c]$ and $\svgauss[h]$
and finally using Eq.~(\ref{eq_svgauss2svint}) also for $\svint(\tsamp)$.
Interestingly, the same reasoning {\em cannot} be made for $\gamext$, i.e. 
it is possible that for quenched configurations with long-ranged correlations
$\gamint=0$ holds but not $\gamext=1/2$.

\section{Models and technical details}
\label{sec_algo}

\subsection{Coarse-grained models}
\label{algo_model}

Various issues discussed theoretically in  Sec.~\ref{sec_theo} will be tested in Sec.~\ref{sec_shear} 
for the fluctuating shear stresses $\tauhat(t)$ measured in computational amorphous solids.
We present numerical results obtained by means of molecular dynamics (MD) and Monte Carlo (MC) simulations 
\cite{AllenTildesleyBook,LandauBinderBook} of three coarse-grained model systems:
\begin{itemize}
\item
quenched elastic networks of repulsive spheres in $d=2$ dimensions connected by harmonic springs.
The networks are created by means of the ``transient self-assem-bled network" (TSANET) model \cite{WKC16,spmP1}
where springs break and recombine locally with an MC hopping frequency $\nu$ changing the connectivity matrix of the network.
The latter MC moves are switched off ($\nu=0$) for all configurations considered in the present work.
Standard MD simulation with a strong Langevin thermostat \cite{AllenTildesleyBook}
moves the particles effectively by overdamped motion through the phase space.
\item
dense polydisperse Lennard-Jones (pLJ) particles in $d=2$ dimensions 
\cite{WTBL02,TWLB02,WXP13,spmP1}.
The configurations are first equilibrated for different temperatures at an imposed average pressure $P=2$
using in addition to standard local MC moves of the particles \cite{LandauBinderBook,WXP13} swap MC moves \cite{Berthier17} 
exchanging pairs of particles. We then switch off the swap MC moves and the barostat. 
Note that each configuration has then a slightly different constant volume $V$.
\item
thin free-standing polmer films suspended parallel to the $(x,y)$-plane \cite{film18,spmP1}
computed by straight-forward MD simulation of a widely used bead-spring model \cite{LAMMPS}.
The films contain $M=768$ monodisperse chains of length $N=16$, i.e. in total $n=12288$ monomers, 
in a periodic box of lateral box size $L=23.5$.
\end{itemize}
A brief presentation of the salient features of each model and the quench protocols
used to create the configurations considered in the present work may be found in Ref.~\cite{spmP1}.

\begin{table*}[t]
\begin{center}
\begin{tabular}{|l|c||c|c|c|}
\hline
property                 & symbol      & TSANET   & pLJ            & films \\ \hline
main simulation method   & -           & MD       & MC             & MD \\
spatial dimension        & $d$         & 2        & 2              & 3 \\
linear simulation box size    & $L$         & 100      & $\approx 103.3$& 23.5 \\
system volume                 & $V$         & $L^2$    & $L^2$          & $L^2H$ \\
particle number               & $n$         & 10000    & 10000          & 12288 \\ 
number density                & $\rho$      & 1        & $\approx 0.94$ & $\approx 1.00$  \\ 
pressure                      & $P$         & 1.7      & 2.0            & $-1.0$ \\ 
temperature                   & $T$         & 1        & 0.2            & 0.05  \\ 
glass transition temperature  & $\Tglass$   & none     & $\approx 0.26$ & $\approx 0.36$  \\ 
number of configurations      & $\Nc$       & 100      & 100            & 100 \\   
maximum sampling time    & $\tsampmax$ & $10^5$   & $10^7$         & $10^5$ \\
measurement time increment& $\tincr$    & 0.01     & 1              & 0.05 \\
plateau of $v(\tsamp)$   & $\vplat$    & 15.3     & 17.1           & $\approx 83$\\
basin relaxation time          & $\taubasin$ & 10       & 2000           & 1 \\
non-ergodicity time            & $\Tnonerg$  & 4200     & 200000         & 800 \\
non-ergodicity parameter       & $\Snonerg$  & 0.16     & 0.25           & 1.13 \\
volume exponent for $\svint$   & $\gamint$ & $\approx 0$   & $\approx 0$   & - \\
volume exponent for $\svext$   & $\gamext$ & $\approx 0.5$ & $\approx 0.44$ & - \\
\hline
\end{tabular}
\vspace*{0.5cm}
\caption[]{Parameters and properties of the models investigated: 
general simulation method,
spatial dimension $d$,
linear size (length) $L$ of periodic simulation box,
system volume $V$,
imposed particle number $n$,
number density $\rho=n/V$,
average normal pressure $P$,
imposed temperature $T$,
glass transition temperature $\Tglass$ for the pLJ particles and the freestanding polymer films,
number of independent configurations $\Nc$,
maximum sampling time $\tsampmax$ for each trajectory,
time increment $\tincr$ between the measured observables,
plateau value $\vplat$ of variance $v(\tsamp)$,
relaxation time of basin $\taubasin$ (Fig.~\ref{fig_v}),
non-ergodicity time $\Tnonerg$ (Fig.~\ref{fig_svtot}),
non-ergodicity parameter $\Snonerg$ (Fig.~\ref{fig_v})
and system-size exponents $\gamint$ and $\gamext$ (Fig.~\ref{fig_V_Snonerg_pdLJ}).
\label{tab}}
\end{center}
\end{table*}

\subsection{Parameters and some properties}
\label{algo_para}

Boltzmann's constant $\kB$, the typical size of the particles (beads) and the particle mass of all models are set to unity
and Lennard-Jones (LJ) units \cite{AllenTildesleyBook} are used throughout this work. 
Time is measured for the pLJ particles in units of MC cycles of the local MC hopping moves of the beads.
Periodic boundary conditions \cite{AllenTildesleyBook,LandauBinderBook} are used for all systems.
The temperature $T$ and the particle number $n$ are imposed. 
Some key properties such as the main simulation method, 
the spatial dimension $d$, 
the linear dimension of the simulation box $L$,
the volume $V$,
the standard particle number $n$,
the working temperature $T$
or the pressure $P$
are summarized in Table~\ref{tab}.\footnote{The
film volume is $V=L^2H$ with $H$ being the film height 
determined from the density profile using a Gibbs dividing surface construction \cite{film18}.
Since the stress tensor vanishes outside the films,
the average vertical normal stress must also vanish for all $z$-planes within the films.
The pressure $P$ indicated for the films in Table~\ref{tab} refers to the normal tangential stresses.}
The number density $\rho = n/V$ is always close to unity.
The working temperature $T$ of the pLJ particles and the polymer films 
are both well below the indicated glass transition temperature $\Tglass$.
(There is no glass transition for the TSANET model.)
The terminal (liquid) relaxation time $\taualpha$ \cite{HansenBook,GraessleyBook} 
of all models is either (by construction) infinite for the quenched elastic networks of the TSANET model 
or many orders of magnitude larger than the maximum sampling time $\tsampmax$ used for the production runs 
of each of the $\Nc$ independent configurations of the ensemble.
The relaxation time $\taubasin$ of the meta-basins may be obtained from the leveling-off of 
$v(\tsamp)$ as shown in Sec.~\ref{shear_v}.
The non-ergodicity parameter $\Snonerg$ is determined equivalently from the large-$\tsamp$ limit of 
$\svtot$ or $\svext$ and $\Tnonerg$ by setting $\svint(\tsamp=\Tnonerg) = \Snonerg$, 
Eq.~(\ref{eq_Tnonerg_def}).
Additional particle numbers $n$ are considered 
for the pLJ particles ($n=100$, $200$, $500$, $1000$, $2000$, $50000$ and $10000$)
in Sec.~\ref{shear_V} where we discuss system-size effects.
We briefly report in Sec.~\ref{shear_T} on preliminary work on temperature effects for the same model
where data for $T=0.19$, $0.2$, $0.25$, $0.3$ and $0.4$ are presented.

\subsection{Observables and data handling}
\label{algo_obs}

The only observable relevant for Sec.~\ref{sec_shear} is the excess contribution $\tauhat$
to the instantaneous shear stress in the $xy$-plane.
See Ref.~\cite{spmP1} for other related properties.
Measurements are performed every $\tincr$ as indicated in Table~\ref{tab}.\footnote{The
standard deviations may depend in addition on the time increment $\tincr$
used to sample the stochastic process \cite{lyuda19a}.
For each model system one unique constant $\tincr$ is thus imposed (cf. Table~\ref{tab}).}
Assuming a pairwise central conservative potential $\sum_l u(\rl)$
with $\rl$ being the distance between a pair of monomers $l$,
the shear stress is given by the off-diagonal 
contribution to the Kirkwood stress tensor \cite{AllenTildesleyBook,TadmorMMBook}
\begin{equation}
\tauhat(t) = \frac{1}{V} \sum_l \rl u^{\prime}(\rl) \ \nlx \nly   
\label{eq_tauexhat}
\end{equation}
with $\nvecl = \rvecl/\rl$ being the normalized distance vector.
The stochastic process $x(t)$ is obtained using Eq.~(\ref{eq_rescale}).
With this rescaling $v[\xbf]$, Eq.~(\ref{eq_vxdef}), characterizes the empirical 
shear-stress fluctuations of the time series and the expectation value $v(\tsamp)$ 
is equivalent to the shear-stress fluctuation modulus $\muF(\tsamp)$ 
considered in previous publications on the stress-fluctuation formalism for elastic moduli
\cite{WXP13,WKC16,ivan17c,ivan18,film18,lyuda19a,spmP1}.
The total standard deviation $\svtot(\tsamp)$ was called $\delta \muF$ in Ref.~\cite{lyuda19a}
and $\delta v$ in Ref.~\cite{spmP1}. 
For clarity we keep below the notations introduced in Sec.~\ref{sec_intro} and Sec.~\ref{sec_theo}.

\label{algo_data}

As indicated in Table~\ref{tab} we prepare for each considered model $\Nc = 100$ independent configurations $c$. 
This allows to probe all properties accurately. For each configuration $c$ we compute and store {\em one} long trajectory 
with $\tsampmax/\tincr \approx 10^7$ data entries. 
Since we want to investigate the dependence of various properties on the sampling time $\tsamp$ 
we probe for each $\tsampmax$-trajectory $\Nk$ equally spaced subintervals $k$ of length 
$\tsamp \le \tsampmax$ with $\Nt=\tsamp/\tincr$ entries. 
Most of the reported results have been obtained for discrete $\Nk$ corresponding to 
$\tsamp = \tsampmax/\Nk$, i.e. $\Nk$ and $\tsamp$ are coupled and all sampled data entries are used 
($\tspacer=0$). As a shorthand we mark these data sets by ``$\Nk \propto 1/\tsamp$".
We remind that $\svint \to 0$ and $\svext \to \svtot$ for $\Nk \to 1$ (Sec.~\ref{theo_ck}).
This limit becomes relevant for $\tsamp \approx \tsampmax$.
We have compared these results with averages taken at fixed constant $\Nk$.
This is done to show that $\svint$ and $\svext$ become rapidly $\Nk$-independent for $\Nk \gg 1$.
Due to the imposed $\tsampmax$ the latter method is limited to $\tsamp \le \tsampmax/\Nk$
and the spacer time interval $\tspacer$ (marked by open circles in Fig.~\ref{fig_intro_sketch}) 
between the sampling time interval $\tsamp$ (filled circles) is not constant but decreases with 
$\Nk$ and $\tsamp$ and vanishes for $\tsamp=\tsampmax/\Nk$. 
Fortunately, the latter point is irrelevant for the non-ergodic systems with 
$\taualph \gg \tsampmax \gg \tsamp + \tspacer \gg \taubasin$, 
i.e. subsequent time series are decorrelated and $\Nk \gg 1$.
It may matter, however, for the analysis of temperature effects as briefly discussed in Sec.~\ref{shear_T}.

\section{Shear-stress fluctuations}
\label{sec_shear}

\subsection{Autocorrelation function $h(t)$}
\label{shear_ht}

\begin{figure}[t]
\centerline{\resizebox{1.0\columnwidth}{!}{\includegraphics*{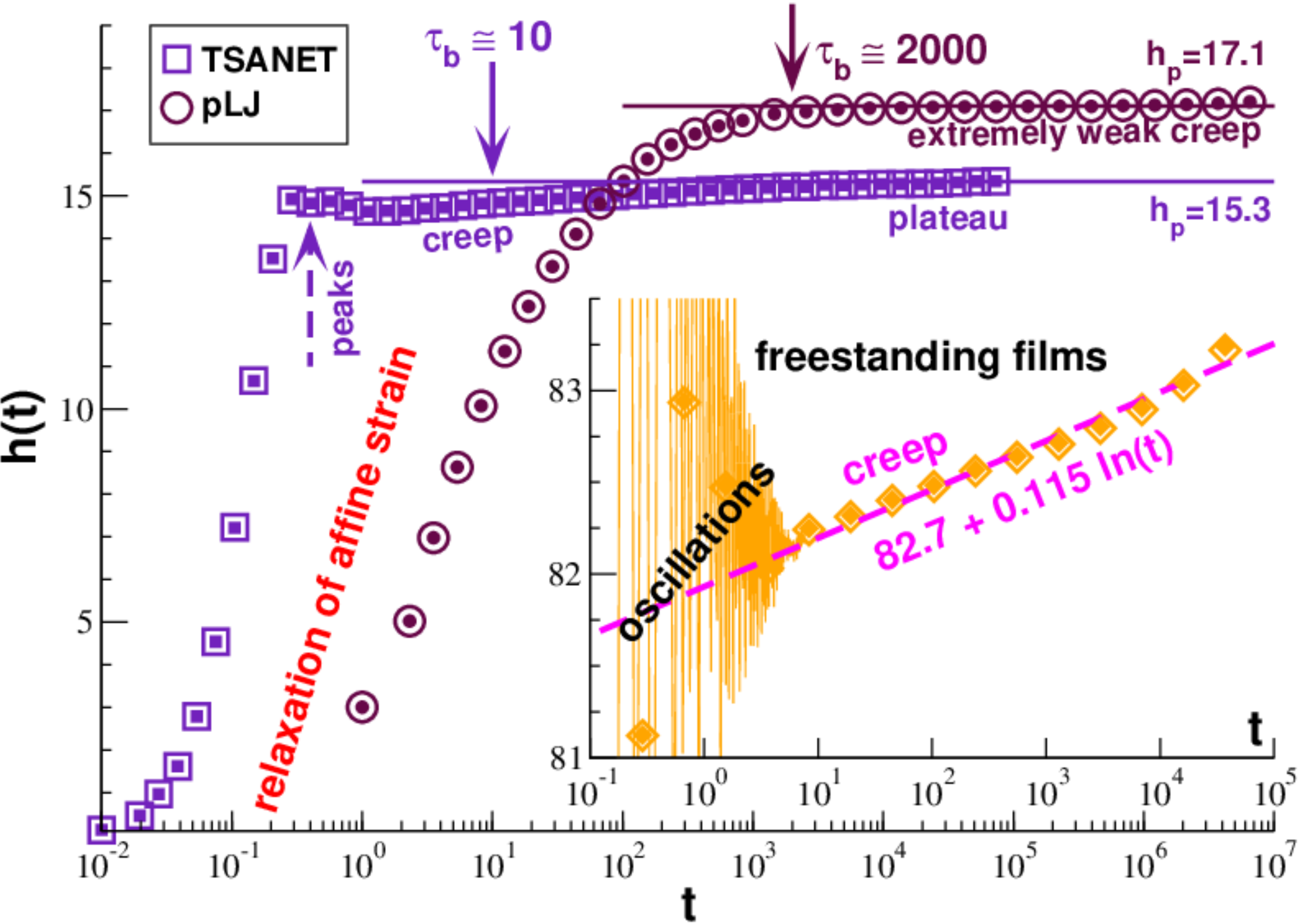}}}
\caption{Shear-stress correlation function $h$ (open symbols) and 
(rescaled) standard deviation $\delta h/\sqrt{2}$ (filled symbols) as functions of time $t$.
The vertical arrows mark the approximate position of $\taubasin$ where $h(t)$ becomes constant.
$\delta h(t)/\sqrt{2} \approx h(t)$ holds to high accuracy confirming the Gaussianity of the stochastic process. 
Inset: Strong short-time oscillations followed by a weak logarithmic creep behavior for polymer films.
}
\label{fig_ht}
\end{figure}

We turn now to the presentation of our numerical results on the shear-stress fluctuations of the three model systems.
As shown in Fig.~\ref{fig_ht} we begin with the ACF $h(t)$. 
We remind that within linear response $h(t)$ is equivalent 
(apart an additive constant $\muA$ and a minus sign)
to the shear-stress relaxation function $G(t)=\muA-h(t)$ \cite{WXP13,WXB15,lyuda19a,spmP1}
commonly measured in experimental studies \cite{FerryBook,GraessleyBook}.
Let us focus first on the data for pLJ particles (circles) obtained by means of local MC moves of the beads
and presented in the main panel. (Time is given for this model in units of MC attempts for all $n$ particles.)
Trivially, $h(0)=0$. $h(t)$ first increases rapidly for $t \ll \taubasin$,
corresponding physically to the relaxation of an affine shear strain imposed at $t=0$ \cite{WXP13,WXB15},
and becomes then essentially constant, $h(t) \to \hplat = 17.1$, for more than three orders of magnitude 
as emphasized by the upper horizontal line. 
To estimate the basin relaxation time $\taubasin \approx 2000$ quantitatively we have used the 
criterion $h(t \approx \taubasin) = f \ \hplat$ setting (slightly arbitrarily) $f=0.99$.
Note that $h(t)$ is strictly monotonically increasing (no oscillations) and that a zoom of the plateau 
regime reveals (not visible) an extremely weak logarithmic creep with $h(t) \approx 16.98+0.01 \ln(t)$ for $t \gg \taubasin$.

The behavior observed for our models using MD simulations (TSANET, polymer films) is unfortunately more complex
revealing both non-monotonic behavior (at short times) and much stronger logarithmic creep.
As may be seen from the main panel, the overdamped TSANET model shows after a maximum at $t \approx 0.3$
(being in fact two peaks superimposed and merged in this representation due to the logarithmic horizontal time scale)
a minimum at $t \approx 1$ followed by a weak logarithmic creep with $h(t) \approx 14.5+0.1 \ln(t)$ up $t \approx 10^4$ 
and then eventually a constant plateau with $\hplat = 15.3$ (middle horizontal line). 
(Using $\tsampmax=10^7$ and $\tincr=1$ we have verified that this is indeed the terminal 
plateau value for these quenched elastic networks.)
What is the relaxation time $\taubasin$ for the meta-basins of the quenched TSANET model?
One reasonable value is $\taubasin \approx 10^4$ characterizing the time where $h(t)$ becomes rigorously constant,
another $\taubasin \approx 10^3$ if we insist on the above criterion with $f=0.99$.
These two values appear, however, far too conservative for many properties discussed below
being integrals over $h(t)$ for which $\taubasin \approx 10$ (vertical arrow) is a more realistic estimate.

The inset presents $h(t)$ for polymer films focusing on the data around $h(t) \approx 82$.
Strong oscillations are seen for short times $t \ll 10$.
The effect is much stronger than for the TSANET model due to the strong bonding potential \cite{film18} 
along the polymer chains and the Nos\'e-Hoover thermostat used for these MD simulations. 
(A strong Langevin thermostat was used for the TSANET model.)
As already pointed out in Ref.~\cite{spmP1}, a logarithmic creep with $h(t) \approx 82.7 + 0.12 \ln(t)$ 
is observed for $t \gg 10$.
The logarithmic creep coefficient is similar to the one observed at intermediate times for the TSANET model
but no final plateau is observed.
The thin polymer films are thus not rigorously non-ergodic, just as the pLJ model.\footnote{Only 
the TSANET systems for $\nu=0$ are rigorously non-ergodic for $\tsampmax \to \infty$. 
The film system is in a transient regime with a wide spectrum of relaxation times both below and above $\tsampmax$. 
As a result Eq.~(\ref{eq_large_tsamp}) cannot hold exactly.
As for the pLJ model, its relaxation time spectrum is apparently well below $\tsampmax$.}
Fortunately, the logarithmic creep coefficients are rather small for all models.
On the logarithmic scales (power-law behavior) we focus on below this effect will be seen to be less crucial
merely causing higher order corrections with respect to the idealized behavior sketched in Sec.~\ref{sec_theo}.

Also indicated in Fig.~\ref{fig_ht} are the rescaled standard deviations $\delta h/\sqrt{2}$ (filled symbols).
As explained in Sec.~III.1 of Ref.~\cite{spmP1}, these were computed using gliding
averages along the trajectories as the last step. We remind that if instantaneous shear stresses 
correspond to a stationary Gaussian process, this implies \cite{spmP1}
\begin{equation}
\delta h(t)^2 = 2 h(t)^2.
\label{eq_dhgauss}
\end{equation}
As can be seen, Eq.~(\ref{eq_dhgauss}) holds nicely for all our models. 
A more precise characterization of the Gaussianity of the stochastic process is obtained using the 
non-Gaussianity parameter $\alpha_2=\delta h^2/2 h(t)^2 -1$ \cite{HansenBook}.
For our standard system sizes this yields very tiny values,
e.g., $\alpha_2 \approx 0.0002$ for pLJ particles.\footnote{The 
non-Gaussianity parameter $\alpha_2$ is seen to increase somewhat for smaller system sizes.
The typical values are, however, always rather small, e.g., $\alpha_2 \ll 0.04$ for all times
for pLJ particles with $n=100$.}

\subsection{Variance $v$ and standard deviation $\svtot$}
\label{shear_v}

\begin{figure}[t]
\centerline{\resizebox{1.0\columnwidth}{!}{\includegraphics*{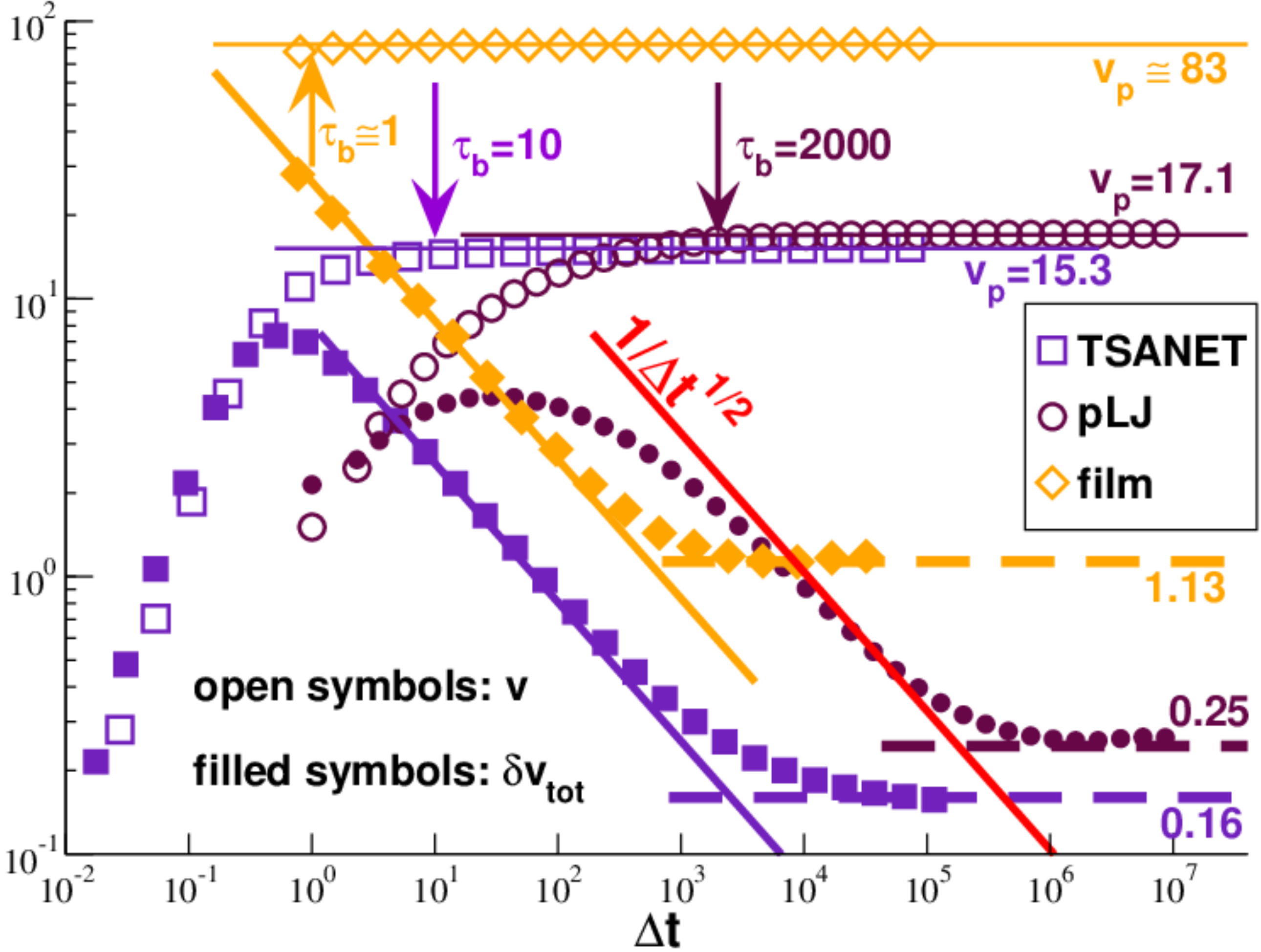}}}
\caption{Shear-stress fluctuation $v$ and the corresponding total standard deviation $\svtot$
(filled symbols) as functions of the sampling time $\tsamp$.
The thin horizontal solid lines mark the long-time plateau value $\vplat$,
the vertical arrows the relaxation time $\taubasin$ of the different models.
While $\svtot \propto 1/\sqrt{\tsamp}$ for intermediate times (bold solid lines),  
a leveling-off $\svtot \to \Snonerg$ is observed for large times (bold dashed horizontal lines)
with $\Snonerg=0.16$ for TSANET, $\Snonerg = 0.25$ for the pLJ particles and
$\Snonerg = 1.13$ for the freestanding polymer films.
}
\label{fig_v}
\end{figure}

Using a double-logarithmic representation we compare in Fig.~\ref{fig_v} 
the shear-stress fluctuation $v$ with the corresponding total standard deviation $\svtot$ (filled symbols). 
We remind that $v(\tsamp)$ is connected with $h(t)$ via Eq.~(\ref{eq_bg_1}).
Being a second integral over $h(t)$, $v(\tsamp)$ is a much smoother and numerically better behaved property \cite{spmP1}.
Due to this $v$ increases monotonically without oscillations and non-monotonic behavior for all three models.
Moreover, since the vertical axis is logarithmic the weak creep of the data mentioned in Sec.~\ref{shear_ht}
becomes irrelevant, i.e. essentially $v(\tsamp) \to \vplat = const$ for $\tsamp \gg \taubasin$ 
as emphasized for all models by the thin horizontal lines marking the plateau value $\vplat$ and 
the vertical arrows for the basin relaxation time $\taubasin$.
(As implied by Eq.~(\ref{eq_bg_1}) $\vplat \approx \hplat$ for all models.)
This allows to definite $\taubasin$ using the {\em same} criterion for all models
by setting
\begin{equation}
v(\tsamp \stackrel{!}{=} \taubasin) = f \ \vplat \mbox{ with } f=0.95
\label{eq_v2taubasin}
\end{equation}
being chosen to obtain the same $\taubasin \approx 2000$ for the pLJ particles as in Sec.~\ref{shear_ht}.
This gives the values stated in Table~\ref{tab}.
(See Fig.~\ref{fig_V_Snonerg_pdLJ} below for the system-size dependence of $\taubasin$ for pLJ particles.)
 
The total standard deviation $\svtot$, computed by averaging over all available time series $\xbf_{ck}$, Eq.~(\ref{eq_dOtot}),
has a maximum about a decade below $\taubasin$.
This is expected from the strong increase of $h(t)$ and $v(\tsamp)$ in this time window \cite{spmP1}. 
As emphasized by the bold solid lines, $\svtot(\tsamp)$ decreases then  
following roughly the $1/\sqrt{\tsamp}$-decay expected for $\taubasin \ll \tsamp \ll \Tnonerg$. 
$\svtot$ becomes constant, $\svtot \to \Snonerg$, for large $\tsamp$ for all models (bold dashed horizontal lines).
As explained in Sec.~\ref{theo_ck}, this is a generic behavior expected for non-ergodic systems.
We determine the values $\Snonerg=0.16$ for TSANET, $\Snonerg = 0.25$ for the pLJ particles
and $\Snonerg = 1.13$ for the freestanding polymer films.
These values are used in the next subsection to rescale the standard deviations $\delta v$.

\subsection{Comparison of $\svgauss$ and $\svint$}
\label{shear_svg_svint}

\begin{figure}[t]
\centerline{\resizebox{1.0\columnwidth}{!}{\includegraphics*{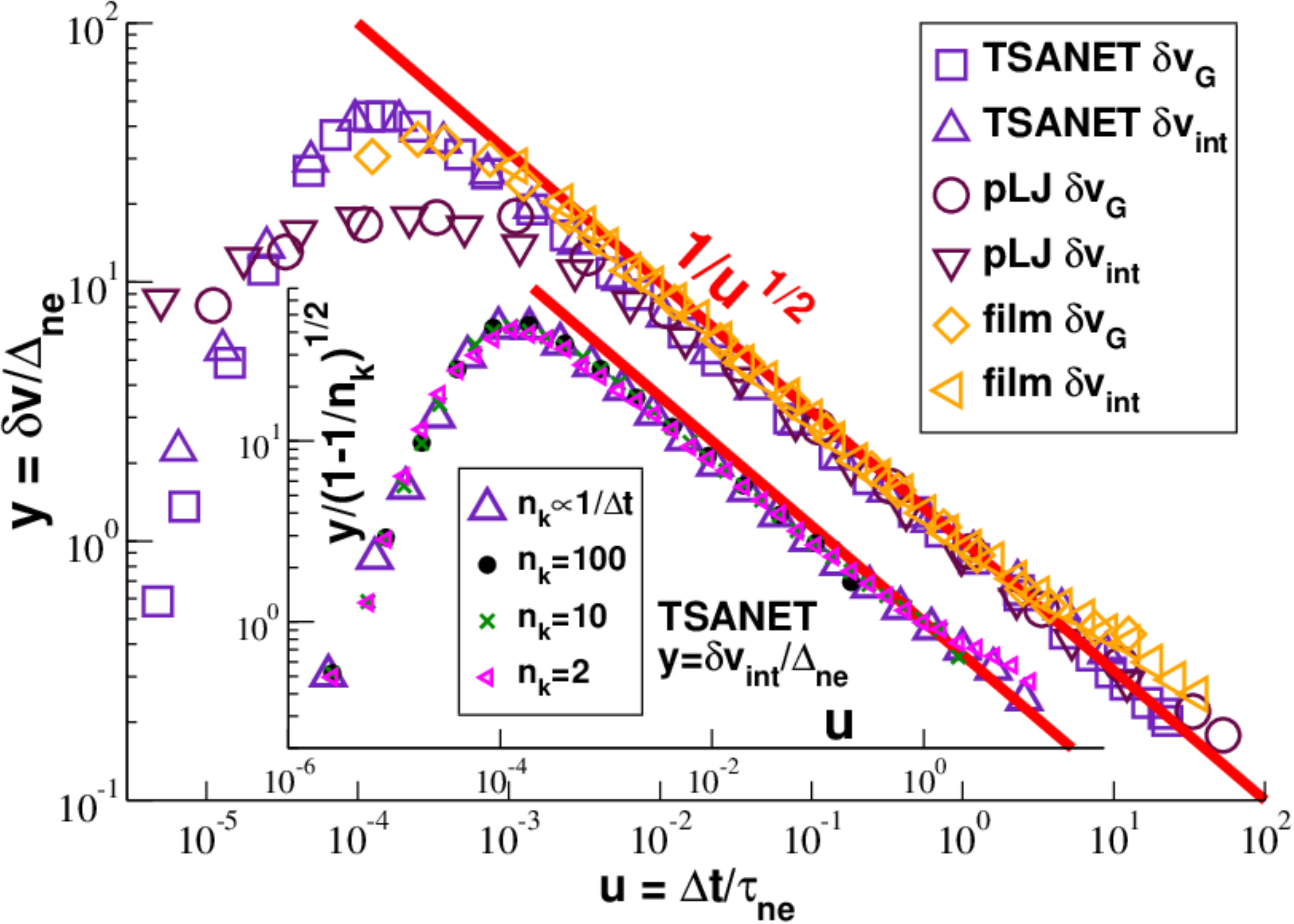}}}
\caption{Main panel: Comparison of $\svgauss$ and $\svint$ using a double-logarithmic representation. 
The reduced standard deviations $y = \delta v/\Snonerg$ are plotted as functions of the reduced 
sampling time $u = \tsamp/\Tnonerg$ with $\Tnonerg=4200$ for the TSANET model,
$\Tnonerg=200000$ for the pLJ particles and $\Tnonerg=800$ for the polymer films. 
The bold solid line marks the expected power-law decay $y \approx 1/\sqrt{u}$.
Inset: 
$y=\svint(\tsamp,\Nk)/\Snonerg$ rescaled as $y/(1-1/\Nk)^{1/2}$ {\em vs.} $u$ for the TSANET model 
and different $\Nk$. The perfect data collapse for $\Nk \ge 2$ is expected from Eq.~(\ref{eq_dv_svint}).
}
\label{fig_svg_svint}
\end{figure}

We compare $\svgauss$ and $\svint$ in the main panel of Fig.~\ref{fig_svg_svint}.
$\svgauss[h]$ has been determined by means of a numerical more suitable reformulation of Eq.~(\ref{eq_bg_2})
described in Refs.~\cite{lyuda19a,spmP1} using the ACF $h(t)$ shown in Fig.~\ref{fig_ht}.
$\svint$ was obtained according to Eq.~(\ref{eq_dOint}) using $\Nk \propto 1/\tsamp$ time series $k$
as described in Sec.~\ref{algo_data}.
Most importantly, $\svgauss \approx \svint$ appears to hold for all $\tsamp$
confirming thus Eq.~(\ref{eq_svgauss2svint}) and the assumption that the
trajectories within each meta-basin are stationary Gaussian processes.
Moreover, plotting the reduced standard deviations $y = \delta v/\Snonerg$ of the three models
as functions of the reduced sampling time $u = \tsamp/\Tnonerg$ leads to a data collapse for 
all three models for $u \gg \taubasin/\Tnonerg$.
Importantly, all data essentially decay as $y \approx 1/\sqrt{u}$ (bold solid line) in the scaling regime.
Note that a free power-law fit would yield a slightly weaker exponent for all models.
This small deviation may be attributed to the fact that the ACFs $h(t)$ of none of the models 
is exactly constant, $h(t) = \hplat$, as shown in Sec.~\ref{shear_ht} at variance to Eq.~(\ref{eq_large_tsamp}).
As already pointed out in Ref.~\cite{spmP1}, deviations are especially seen for polymer films for $u \gg 1$.

The inset of Fig.~\ref{fig_svg_svint} presents in more detail $y(u) = \svint/\Snonerg$ for the TSANET model
comparing data obtained for different numbers $\Nk$ of time series $k$ for each configuration $c$.
The large triangles represent the same data shown in the main panel where $\Nk \propto 1/\tsamp$,
all other data have been obtained with a fixed number $\Nk$ as indicated.
We remind that $\svint=0$ for $\Nk=1$. 
A direct plot of $y$ (not shown) reveals that all data but those for $\Nk \le 10$ collapse,
i.e. the $\Nk$-dependence becomes rapidly irrelevant. An even better data collapse for all data 
with $\Nk \ge 2$ is obtained as suggested by Eq.~(\ref{eq_dv_svint}) using the rescaled  
standard deviation $y/(1-1/\Nk)^{1/2}$. 
In other words it is sufficient to use $\Nk=2$ time series for one configuration to obtain using
the rescaling factor $(1-1/\Nk)^{1/2}$ the asymptotic limit. 
This finding should strongly simplify future numerical work.

\subsection{Comparison of $\svint$ and $\svtot$}
\label{shear_svtot}

\begin{figure}[t]
\centerline{\resizebox{1.0\columnwidth}{!}{\includegraphics*{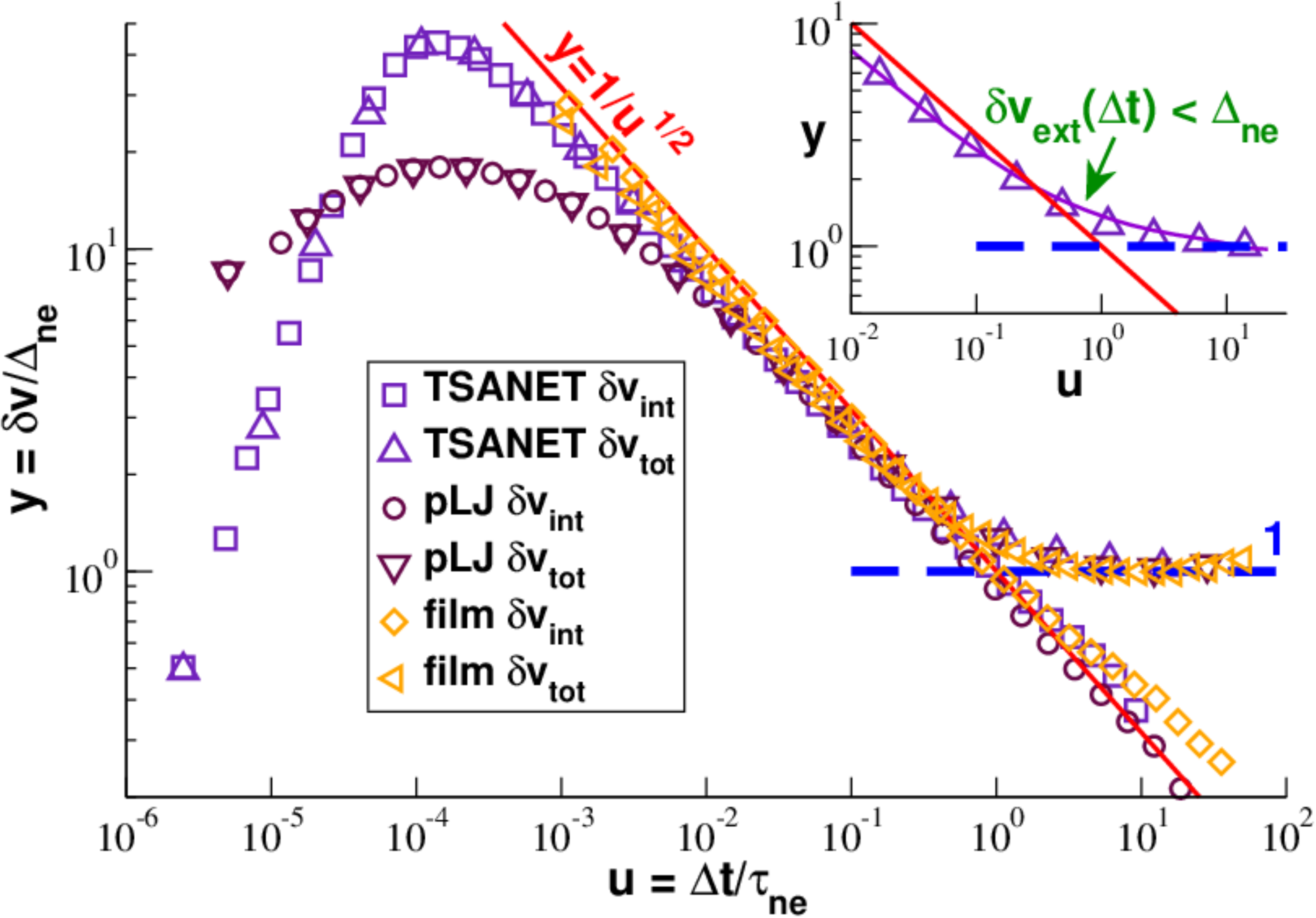}}}
\caption{Main panel:
Comparison of $\svint$ and $\svtot$ using double-logarithmic coordinates with 
$y=\delta v/\Snonerg$ and $u=\tsamp/\Tnonerg$. 
$\svint \approx \svtot$ holds for $u \ll 1$ while $\svtot \to 1$ for $u \gg 1$ (bold dashed line).
Inset: $y=\svtot(\tsamp)/\Snonerg$ vs. $u$.
As shown for the TSANET model, Eq.~(\ref{eq_svtot_intpol}) gives a good approximation for $\svtot$.
Tiny deviations are seen for $u \approx 1$.
}
\label{fig_svtot}
\end{figure}

We compare $\svtot$ with $\svint$ in Fig.~\ref{fig_svtot} using reduced units with 
$y=\delta v/\Snonerg$ and $u=\tsamp/\Tnonerg$.
We remind that $\Tnonerg$ (Tab.~\ref{tab}) has been determined as a crossover time 
by means of Eq.~(\ref{eq_Tnonerg_def}) 
using the measured $\Snonerg$ and $\svint(\tsamp)$. Apart very short (reduced) sampling times $u$,
the rescaled data depend very little on the model on the logarithmic scales considered. 
As expected, $\svtot \approx \svint$ holds to high precision for all $u \ll 1$.
All data sets decrease essentially as $y \approx 1/\sqrt{u}$ for $u \gg \ubasin$
over nearly three orders of magnitude as emphasized by the bold solid line.
While the $1/\sqrt{u}$-decay continues for $\svint$ for large $u \gg 1$, 
the rescaled $\svtot$-data levels off to the plateau indicated by the horizontal dashed line.
  
Focusing on the TSANET model we test the interpolation formula Eq.~(\ref{eq_svtot_intpol}) 
in the inset of Fig.~\ref{fig_svtot}, i.e. we compare the directly measured $\svtot$ (triangles) 
with $(\dvint(\tsamp)+\Snonerg^2)^{1/2}$ (solid line).\footnote{Eq.~(\ref{eq_svtot_intpol}) 
is applicable for $\tsamp \gg \taubasin$. In terms of $u$ this condition becomes
$u \gg 1/400$ for the TSANET model. This is roughly satisfied by the $u$-range
presented in Fig.~\ref{fig_svtot}.}
The same result is obtained by replacing $\svint$ by $\svgauss$ as expected from Fig.~\ref{fig_svg_svint} (not shown).
The interpolation formula is seen to give an excellent fit of $\svtot$.
To leading order $\svtot$ is thus given by $\svint \approx \svgauss$ and, hence, by $h(t)$ {\em plus} an additional constant.
As indicated by the arrow, Eq.~(\ref{eq_svtot_intpol}) slightly {\em overpredicts} $\svtot$ for $u \approx 1$. 
Apparently, $\svext(u)$ approaches its asymptotic limit $\Snonerg$ from below.

\subsection{Characterization of $\svext(\tsamp,\Nk)$}
\label{shear_svext}

\begin{figure}[t]
\centerline{\resizebox{1.0\columnwidth}{!}{\includegraphics*{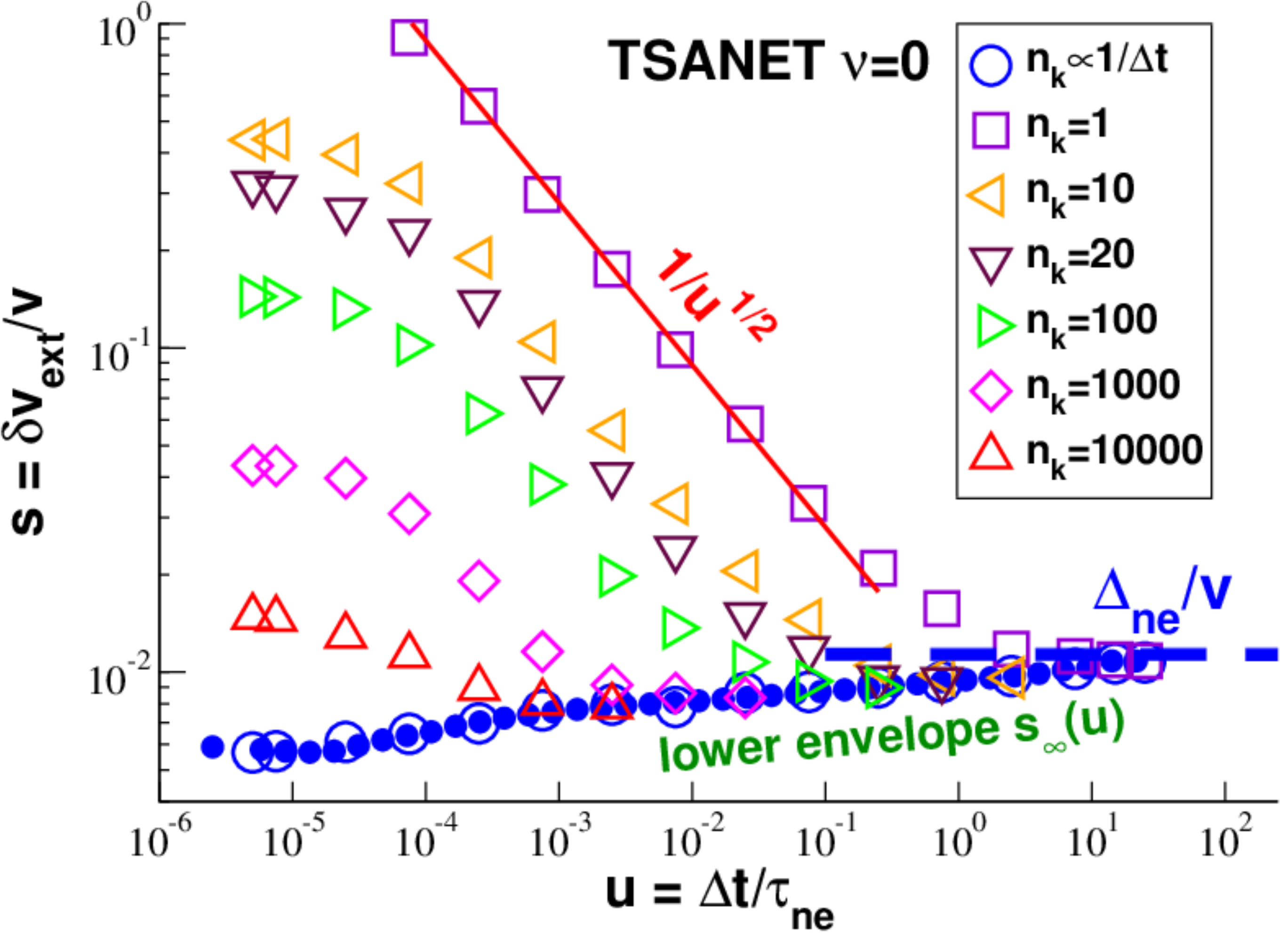}}}
\caption{$s = \svext/v$ {\em vs.} $u=\tsamp/\Tnonerg$ for quenched TSANET networks.
The large open and small filled circles have been obtained using $\Nk \propto 1/\tsamp$,
all other symbols by imposing a constant $\Nk$.
The thin solid line indicates the expected power-law behavior $s \approx 1/\sqrt{u}$
for small $\Nk$, the bold dashed horizontal line the asymptotic limit
$s \to \Snonerg/v$ for $u \gg 1$.
}
\label{fig_svext_TSANET}
\end{figure}

This point is further investigated in Fig.~\ref{fig_svext_TSANET}
presenting the dimensionless standard deviation 
$s = \svext(\tsamp,\Nk)/v(\tsamp)$ for the TSANET model.
(See Fig.~\ref{fig_V_tsamp_pdLJ} for the unscaled $\svext$-data for pLJ particles.)
As emphasized in Sec.~\ref{theo_ck}, $\svext$ depends in general on $\tsamp$ and may also depend on $\Nk$.
The data indicated by the large open and the small filled circles
have been both obtained for $\Nk \propto 1/\tsamp$ as described in Sec.~\ref{algo_data}. 
To demonstrate the {\em numerical} equivalence of both definitions 
the small filled circles are computed using $\dvext=\dvtot -\dvint$, Eq.~(\ref{eq_key_1}),
and the large circles using directly Eq.~(\ref{eq_dOext}).
 
It is also instructive to characterize $s$ for different fixed numbers $\Nk$ of 
equidistant and non-overlapping time series decoupling thus $\tsamp$ and $\Nk$. 
We remind that $\svext = \svtot$ for $\Nk=1$ and the power-law slope indicated for the
intermediate $\tsamp$-regime of this data set corresponds to the $1/\sqrt{\tsamp}$-decay
already shown in Fig.~\ref{fig_svtot}.
Confirming Sec.~\ref{theo_ck}, $s$ becomes $\Nk$-independent for large $\Nk$ 
approaching a lower envelope $\sinf(\tsamp) = \lim_{\Nk \to \infty} s(\tsamp,\Nk)$ from above.
This lower envelope corresponds essentially to the circles. 
$\sinf(\tsamp)$ is seen to {\em increase monotonically}, albeit extremely weakly,
approaching $\Snonerg/v$ (dashed line) from below. This is consistent with the tiny
deviations from $\svtot$ observed for the shifted $\svint$-data in the inset of Fig.~\ref{fig_svtot}. 
Similar results have been obtained for the other models as seen in the inset of 
Fig.~\ref{fig_svext_pdLJ} showing $\svext(\tsamp,\Nk)$ for the pLJ particles.

We note finally that Eq.~(\ref{eq_dOext_Nk_large_tsamp}) 
implies in principle that
\begin{equation}
\Nk \left( \dvext(\tsamp,\Nk)-\Dnonerg \right) \simeq \dvint(\tsamp) \ge 0
\label{eq_dvext_Nk_problem}
\end{equation}
for $\tsamp \gg \taubasin$. This allows to express $\svext(\tsamp,\Nk)$
in terms of $\svint(\tsamp)\approx \svgauss[h]$ for small $\tsamp$ and $\Nk$ (not shown).
Unfortunately, this is not possible in the opposite limit since $\dvext(\tsamp,\Nk)-\Dnonerg$ 
becomes negative as seen by the monotonic increase of $\sinf(\tsamp)$.
It is better to go back to the more general Eq.~(\ref{eq_dOext_Nk}) which can be
rephrased as
\begin{equation}
\svext(\tsamp) \simeq (\dvext(\tsamp,\Nk) - \dvgauss[h]/\Nk)^{1/2}.
\label{eq_dvext_Nk_problem2}
\end{equation}
As shown in the inset of Fig.~\ref{fig_svext_pdLJ} by the large crosses for $\Nk=10$
this may be used to obtain the asymptotic $\svext(\tsamp)$ from $\svext(\tsamp,\Nk)$
and $\svgauss[h]$, at least if $\svgauss[h]$ is available with sufficient precision.

\subsection{Temperature dependence of $\svext$}
\label{shear_T}

\begin{figure}[t]
\centerline{\resizebox{.95\columnwidth}{!}{\includegraphics*{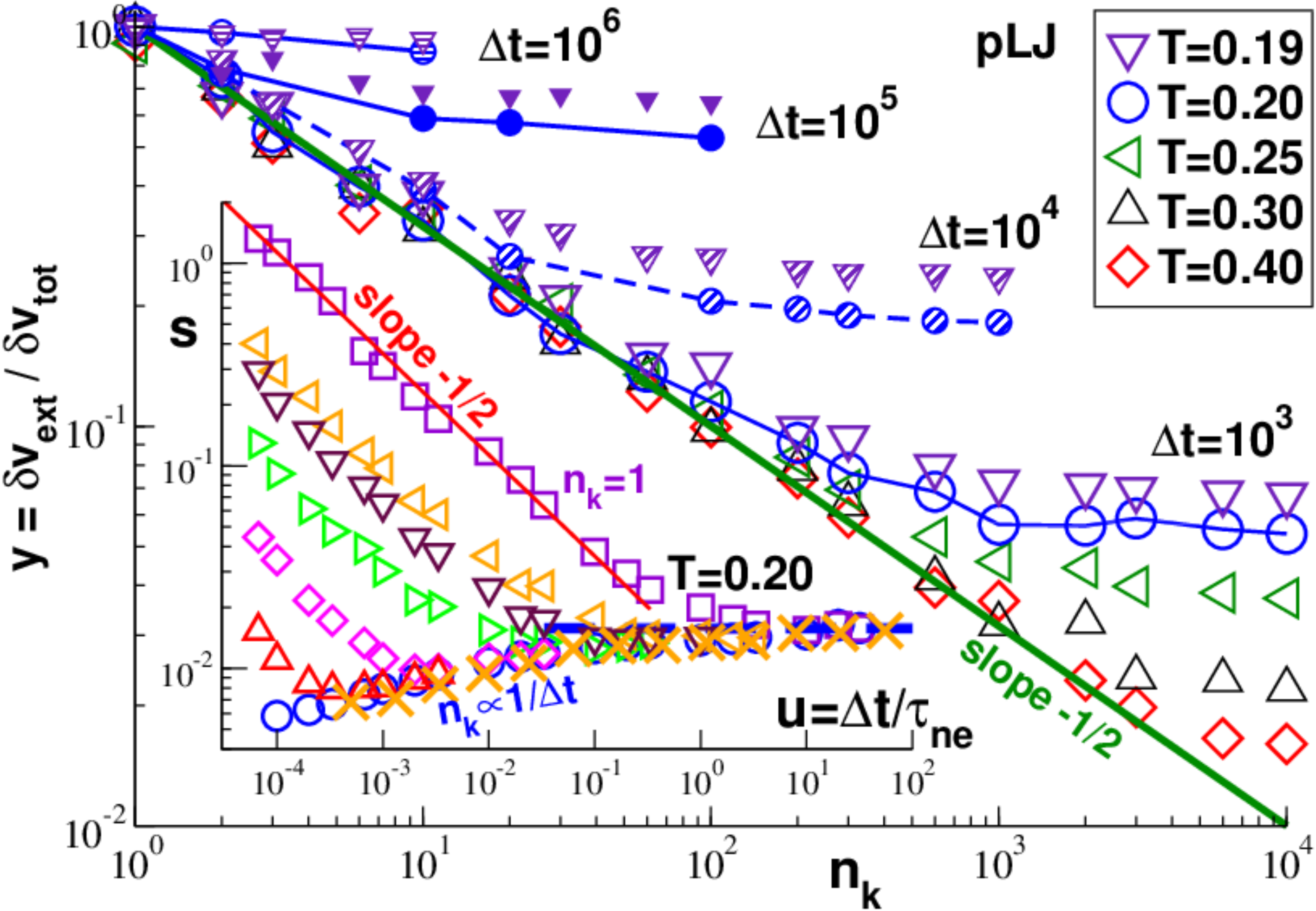}}}
\caption{$\svext(\tsamp,\Nk)$ for pLJ particles.
Inset:
$s = \svext/v$ {\em vs.} $\tsamp$ for $T=0.2$ for different $\Nk$ 
using the same symbols as in Fig~\ref{fig_svext_TSANET}. 
The thin solid line indicates the power-law slope $-1/2$ for $\svext(\Nk=1)=\svtot$, 
the solid dashed line the large-$\tsamp$ limit $\Snonerg/v$.
With increasing $\Nk$ all data sets approach a lower $\Nk$-independent envelope $\sinf(\tsamp)$.
A good estimation of this limit is given by the data for $\Nk \propto 1/\tsamp$ (circles).
The crosses represent the rescaled $\svext(\tsamp,\Nk)$ for $\Nk=10$ using $\svgauss[h]$
and the approximation Eq.~(\ref{eq_dvext_Nk_problem2}).
Main panel:
$y = \svext/\svtot$ {\em vs.} $\Nk$ for $\tsamp=10^3$, $10^4$, $10^5$ and $10^6$ (from bottom to top)
for several temperatures. Open symbols are used for $\tsamp=10^3$, filled symbols for $\tsamp=10^5$.
The data for $T=0.2$ is connected by lines.
The bold solid line indicates the power-law $-1/2$ expected for independent time series. 
}
\label{fig_svext_pdLJ}
\end{figure}

A different representation of $\svext$ is chosen in the main panel of Fig.~\ref{fig_svext_pdLJ} 
where data sets for fixed sampling times $\tsamp$ (increasing from bottom to top) are plotted as functions of $\Nk$.
Extending beyond the main focus of this work on non-ergodic systems we compare here data sets for a broad range of temperatures $T$. 
The dimensionless vertical axis $y = \svext(\Nk)/\svext(\Nk=1)$ is used to normalize all data sets
for different $\tsamp$ and $T$ and to compare $\svext$ with $\svtot=\svext(\Nk=1)$.
$y \ll 1$ implies that $\svtot \approx \svint$, i.e. both averaging procedures become equivalent.
The bold solid line indicates the power law $1/\sqrt{\Nk}$ expected for independent time series $\xbf_{ck}$ 
being a {\em lower envelope for all data sets}. This envelope is the more relevant the smaller $\tsamp$ and the higher $T$.
This is especially the case for all high temperatures where the systems are ergodic
and according to Eq.~(\ref{eq_dOext_ergodic}) we have
\begin{equation}
\svext(\tsamp,\Nk) \simeq \frac{\svint(\tsamp)}{\sqrt{\Nk}} = \frac{\svgauss[h]}{\sqrt{\Nk}}
\label{eq_svext_ergodic}
\end{equation}
for $\Nc \to \infty$.
In agreement with Fig.~\ref{fig_svext_TSANET} and the inset of Fig.~\ref{fig_svext_pdLJ},
$\svext$ increases with $\tsamp$ and becomes $\Nk$-independent for large $\tsamp$ and low $T$.
Note that the $\Nk$-dependence is weak for $\tsamp=10^6$ and $T=0.2$ and $T=0.19$.

A technical issue relevant for future work should be mentioned here.
Closer inspection of the data for $T=0.3$ and $T=0.4$ shows in fact a small upbending for the largest $\Nk$
which is not consistent with Eq.~(\ref{eq_svext_ergodic}).
We remind that we have stored for each configuration $c$ only one trajectory of constant length $\tsampmax$, 
i.e. the spacer interval $\tspacer$ between the used time series of length $\tsamp \le \tsampmax/\Nk$ decreases 
strongly with $\tsamp$ and $\Nk$. Neighboring $\tsamp$-intervals become thus correlated once $\tspacer$ gets smaller 
than the terminal relaxation time $\taualpha(T)$ \cite{HansenBook,GraessleyBook}.  
One simple means to test that the observed upbending at high temperatures is merely due to this technical effect
would be to increase $\tsampmax$ and thus $\tspacer$ by, say, a factor $10$ or $100$. The upbending must then be
shifted to correspondingly larger $\tsamp$. Larger $\tsampmax$ are in any case warranted to better show for 
$T \ll \Tglass$ that $\svext(\tsamp) \to \Snonerg$ for large $\tsamp$.
However, for a physical meaningful characterization of $\svext$ for intermediate temperatures 
it would be even better to work with a {\em constant} spacer time $\tspacer$
for all temperatures and to sample thus $\Nk$ sequences of fixed spacer and measurement time intervals
decoupling thus $\Nk$ from both $\tsamp$ and $\tspacer$. 
$y \approx 1/\sqrt{\Nk}$ must then rigorously hold for $\tspacer \gg \taualpha$ 
while $y$ should reveal a (possibly temperature dependent) shoulder in the opposite limit.
The next challenge to be addressed then is of whether a time-temperature superposition scaling using the directly measured 
terminal relaxation $\taualpha(T)$ is possible or not.

\subsection{System-size dependence}
\label{shear_V}

\begin{figure}[t]
\centerline{\resizebox{1.0\columnwidth}{!}{\includegraphics*{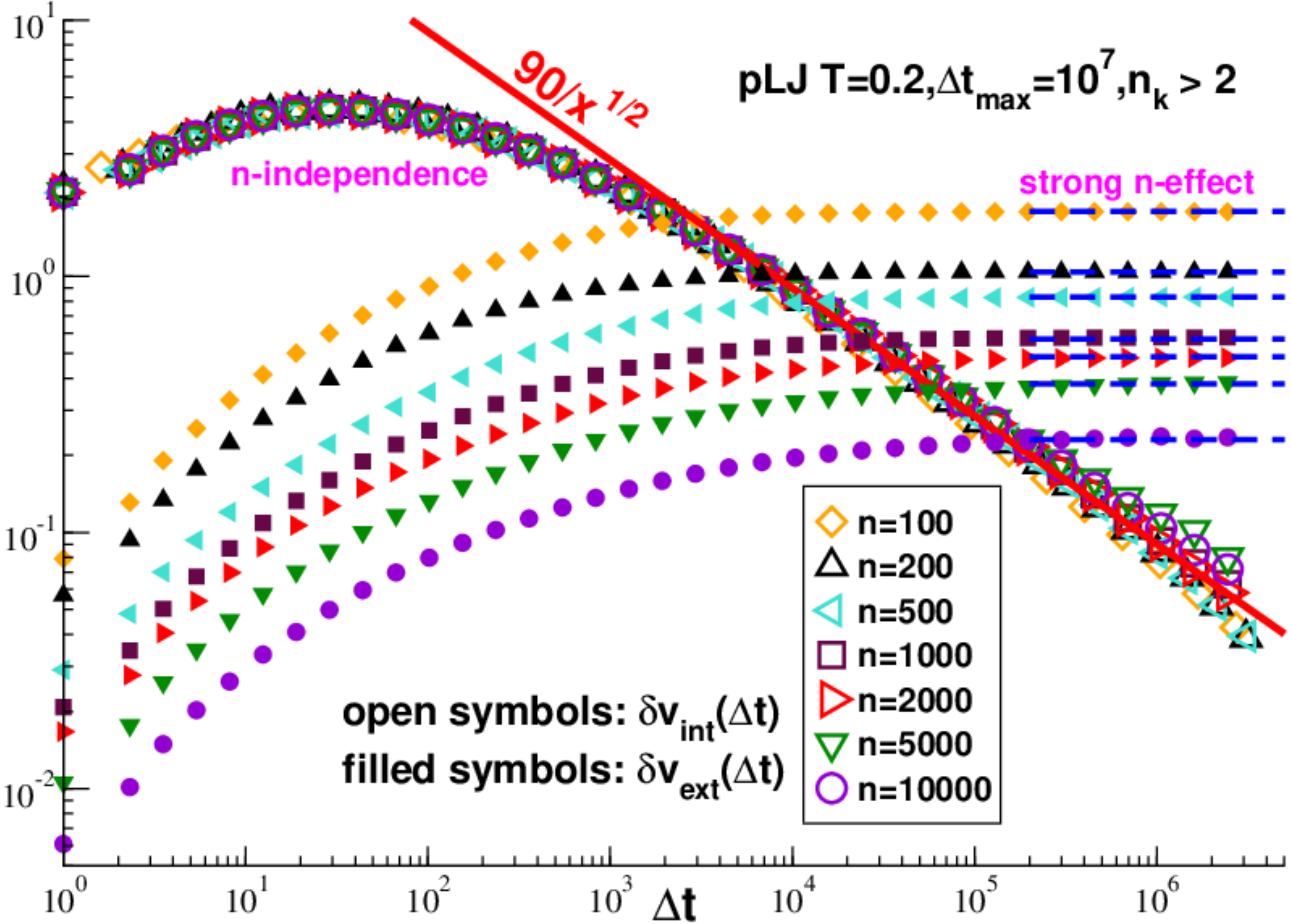}}}
\caption{$\svint$ (open symbols) and $\svext$ for pLJ systems for a broad range of particle numbers $n$.
$\svint$ is essentially $n$-independent, i.e. $\gamint=0$, while $\svext$ decreases with $n$.
The bold solid line indicates the decay of $\svint$ expected according to Eq.~(\ref{eq_large_tsamp}),
the dashed horizontal lines show the $\svext$-values given in Fig.~\ref{fig_V_Snonerg_pdLJ}.
}
\label{fig_V_tsamp_pdLJ}
\end{figure}

We investigate now the dependence of several properties on the system size
focusing on data obtained for the pLJ particles.
We have seen above that the total variance $\dvtot$ of the shear-stress fluctuations $v$
of quenched elastic bodies may be decomposed as the sum of two contributions due to independent physical 
causes: the internal and external variances $\dvint$ and $\dvext$.
The main point made in this subsection is that $\svint$ and $\svext$ are characterized by different 
$n$-dependences.
Figure~\ref{fig_V_tsamp_pdLJ} compares the $\tsamp$-dependences of $\svint$ and $\svext$ for different particle numbers $n$. 
$\svint$ and $\svext$ have been computed using $\Nk \propto 1/\tsamp$ time series for each configuration.
The data are plotted as functions of the unscaled sampling time $\tsamp$ in units of MC steps.
The bold solid line indicates the decay of $\svint$ expected according to Eq.~(\ref{eq_large_tsamp}) 
for $\tsamp \gg \taubasin$.
As can be seen, $\svint$ is essentially $n$-independent, 
i.e. $\gamint=0$ as expected if standard statistical physics holds for each meta-basin. 
At striking variance to this $\svext$ strongly decreases with $n$, i.e. the $v_c$ become similar,
and becomes constant, $\svext \to \Snonerg$, for large $\tsamp$.
Interestingly, $\svint(\tsamp)$ is a monotonically decreasing function of $\tsamp$
while $\svext(\tsamp)$ is always monotonically increasing. 
Note that the increase of $\svext(\tsamp)$ for $\tsamp \ll \taubasin$ is much stronger than
the one seen for the reduced external standard deviation $\sinf(\tsamp)$ in Fig.~\ref{fig_svext_TSANET} 
and the inset of Fig.~\ref{fig_svext_pdLJ}.
In other words, the $\tsamp$-dependence of $\svext(\tsamp)$ stems mainly from the $\tsamp$-dependence of $v(\tsamp)$, Fig.~\ref{fig_v}.

\begin{figure}[t]
\centerline{\resizebox{1.0\columnwidth}{!}{\includegraphics*{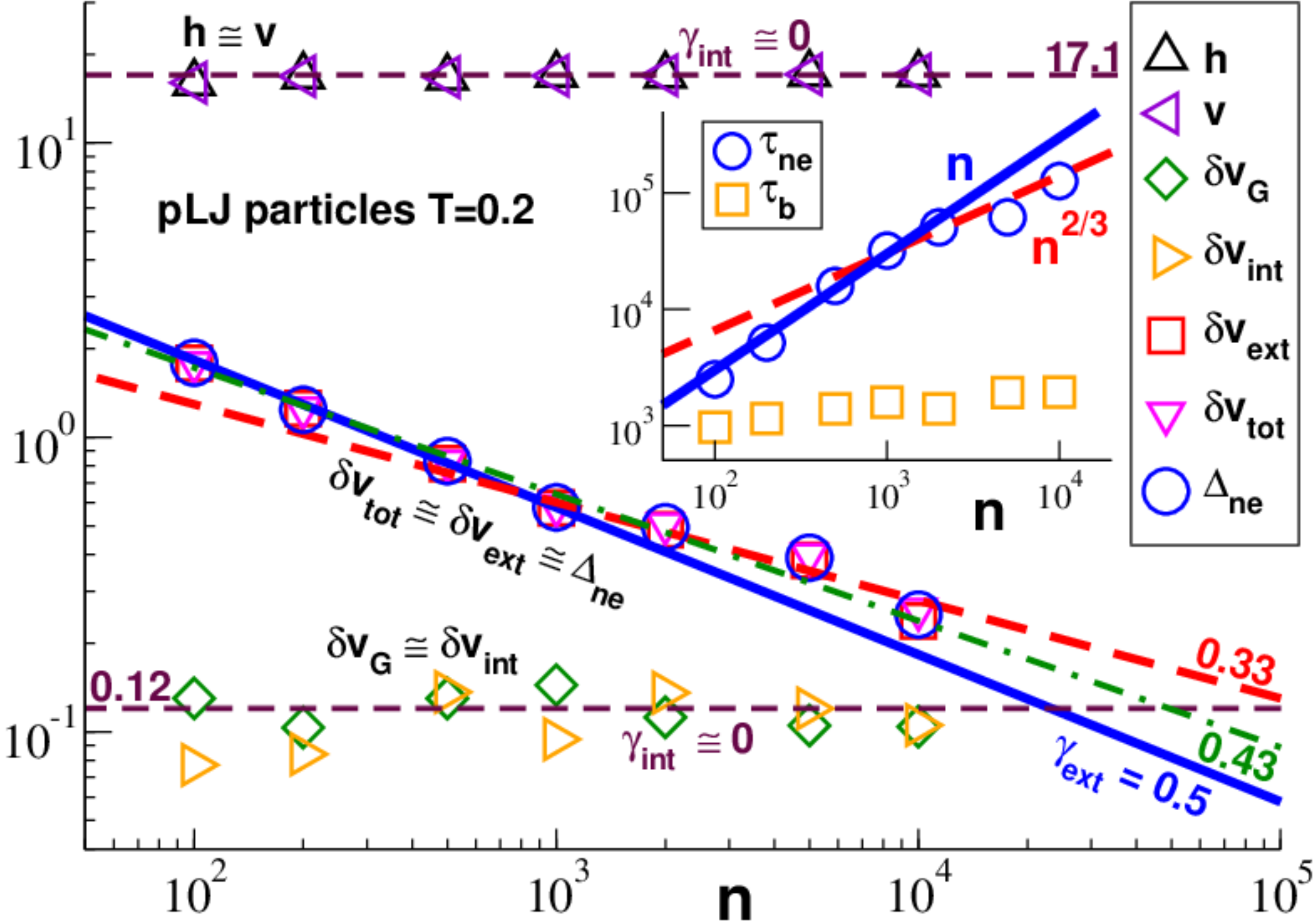}}}
\caption{Dependence on particle number $n$ for various properties for pLJ particles.
Main panel: 
$h$, $v$, $\svgauss$, $\svint$, $\svtot$ for $\tsamp=10^6$ and $\Snonerg$.
The thin horizontal dashed lines indicate the exponent $\gamint=0$,
the bold solid line $\gamext = 1/2$,
the dash-dotted line $\gamext \approx 0.43$ and the bold dashed line $\gamext \approx 0.33$.
Inset:
$n$ dependence of $\taubasin$ and $\Tnonerg$. 
While $\taubasin$ saturates for large $n$, $\Tnonerg$ increases with $n$ 
broadly in agreement with Eq.~(\ref{eq_Tnonerg_V}).
}
\label{fig_V_Snonerg_pdLJ}
\end{figure}


The $n$-dependence of various properties is presented in Fig.~\ref{fig_V_Snonerg_pdLJ}. 
We compare in the main panel $h$, $v$, $\svgauss$, $\svint$, $\svext$ and $\svtot$ measured at $\tsamp=10^6$ 
with the non-ergodicity parameter $\Snonerg$ (circles).
As emphasized by the dashed horizontal lines, 
$h$, $v$ and $\svgauss \approx \svint$ are all independent of the particle number $n$,
i.e. $\gamint=0$ as expected from Sec.~\ref{theo_T}.
Moreover, $\svext$, $\svtot$ and $\Snonerg$ are within numerical accuracy identical.
This is expected since $\tsamp=10^6 \gg \Tnonerg$ for all $n$. 
$\Snonerg$ was seen to decrease with a power-law exponent $\gamext = 1/2$ for the TSANET model \cite{spmP1}.
According to Eq.~(\ref{eq_gamma_uncorr}) this suggests that independent localized shear-stress 
fluctuations are relevant for these elastic networks. 
Interestingly, a weaker exponent $\gamext \approx 1/3$ (bold solid line)
has been fitted in recent simulation studies of 2D binary LJ mixtures \cite{Procaccia16},
dense 3D polymer glasses \cite{lyuda19a} and to the 2D pLJ particles \cite{spmP1}
also investigated in the present study.
A somewhat larger exponent $\gamext \approx 0.43$ (dash-dotted line)
appears to better fit all currently available pLJ data.
Assuming that future simulations confirm that $\gamext < 1/2$ this could be
explained by long-range spatial correlations with a {\em diverging} correlation length $\xi$ 
\cite{lyuda18,lyuda19a,Gardner}.
 
As can be seen from the inset, the basin relaxation time $\taubasin$,
obtained using Eq.~(\ref{eq_v2taubasin}) from $v(\tsamp)$, only depends weakly (logarithmically) on $n$.
At variance to this $\Tnonerg(n)$, obtained using Eq.~(\ref{eq_Tnonerg_def}), strongly increases.
The two indicated power-law slopes are attempts to characterize this dependence.
According to Eq.~(\ref{eq_Tnonerg_V}) one expects 
$\Tnonerg \propto n^{2\gamext}$ for $\gamint \approx 0$.
Depending on whether $\gamext=1/2$ or $\gamext \approx 1/3$, 
this corresponds either to $\Tnonerg \propto n$ (bold solid line) 
or $\Tnonerg \propto n^{2/3}$ (dashed line). 
The linear relation only fails for the two largest systems.
 
\section{Conclusion}
\label{sec_conc}

Extending our recent work focusing on ergodic stationary Gaussian stochastic processes \cite{lyuda19a,spmP1} 
on to non-ergodic systems,
we have described in general terms the standard deviation $\delta v(\tsamp)$ of the empirical variance $v[\xbf]$, 
Eq.~(\ref{eq_vxdef}), of time series $\xbf$ measured over a finite sampling time $\tsamp$.
Since independent ``configurations" $c$ get trapped in meta-basins of the generalized phase space 
(Fig.~\ref{fig_intro_sketch}) 
it becomes relevant in which order $c$-averages and $c$-variances over configurations $c$
and $k$-averages and $k$-variances over time series $k$ of a given configuration $c$ (Sec.~\ref{theo_definitions})
are performed.
Three types of variances of $v[\xbf_{ck}]$ must be distinguished: 
the total variance $\dvtot$, Eq.~(\ref{eq_dOtot}),  
the internal variance $\dvint$ within each meta-basin, Eq.~(\ref{eq_dOint}),
and the external variance $\dvext$ between the different basins, Eq.~(\ref{eq_dOext}).
It was shown (Sec.~\ref{theo_ck}) that $\dvtot = \dvint + \dvext$, 
Eq.~(\ref{eq_key_1}).
Various general and more specific simplifications of our key relation Eq.~(\ref{eq_key_1}) 
are given for physical systems where the stochastic process $x(t)$ is due to a fluctuating
density field averaged over the system volume $V$.
Assuming the stochastic process within each basin to be thus (essentially) Gaussian, $\svint$ is given by
the functional $\svgauss[h]$, Eq.~(\ref{eq_bg_2}), in terms of the $c$-averaged ACF $h$, 
Eq.~(\ref{eq_svgauss2svint}). Both the $\tsamp$- and the $V$-dependence of $\svint$ is thus imposed by $h(t)$.
Specifically, this implies that $\svint(\tsamp) \approx \svgauss(\tsamp) \propto 1/\sqrt{\tsamp}$ for $\tsamp \gg \taubasin$.
Moreover, $\svext$ converges for $\tsamp \gg \taubasin$ to the constant ``non-ergodicity parameter" $\Snonerg$. 
Since $\svext \approx \Snonerg$ decreases more strongly with the system volume $V$ than $\svint$ (Sec.~\ref{theo_T}),
the non-ergodicity time $\Tnonerg(V)$, Eq.~(\ref{eq_Tnonerg_def}), must increase with $V$.
Deviations of $\svtot$ from $\svint \approx \svgauss$ are thus merely finite-size effects.

We have illustrated and essentially confirmed these relations in Sec.~\ref{sec_shear}
for stochastic processes obtained from the (reduced) shear stresses $x(t) = \sqrt{\beta V} \tauhat(t)$
computed in amorphous solids. Quenched elastic networks and two low-temperature glasses have been compared.
The Gaussianity approximation $\svint \approx \svgauss[h]$, Eq.~(\ref{eq_svgauss2svint}), is seen to hold 
for all $\tsamp$ (Fig.~\ref{fig_svg_svint}), i.e. $\svint(\tsamp)$ is set by $h(t)$.
Interestingly, $\svext$ is seen to approach its asymptotic limit $\svext \approx \Snonerg$ 
from below (Figs.~\ref{fig_svext_TSANET}, \ref{fig_svext_pdLJ} and \ref{fig_V_tsamp_pdLJ}).
The discussion in Secs.~\ref{shear_svg_svint}-\ref{shear_svext} 
has focused on the comparison of $\svint$, $\svgauss$, $\svtot$ and $\svext$
for one state point, i.e. one temperature and one system size. 
Effects of the volume $V \propto n$ have been considered in Sec.~\ref{shear_V}. 
While $h$, $v$, $\svgauss \approx \svint$ are essentially $V$-independent ($\gamint \approx 0$)
as expected for stochastic processes of intensive thermodynamic fields (Sec.~\ref{theo_T}),
$\svext \approx \Snonerg \propto 1/V^{\gamext}$ strongly decreases (Fig.~\ref{fig_V_Snonerg_pdLJ}). 
That $\svint$ and $\svext$ are independent contributions to $\svtot$
characterized by different statistics is thus manifested by their different $V$-dependences.
While an exponent $\gamext=1/2$ has been fitted for the TSANET model \cite{spmP1},
a weaker (apparent) exponent $\gamext < 1/2$ appears to fit $\Snonerg$ for the pLJ particles.
As already pointed out elsewhere \cite{lyuda19a} this suggests long-range spatial correlations.

Temperature effects have been mentioned briefly for pLJ particles and the external variance $\dvext$ (Sec.~\ref{shear_T}).
As pointed out there, future studies should increase the total sampling times
$\tsampmax$ for each configuration to better describe the scaling of $\svint$ and $\svext$
with $\tsamp$ and $\Nk$ for different temperatures. Especially, it should be useful to sample
these properties using a fixed spacer time interval $\tspacer$ for all temperatures. 
While $\svext(\Nk) \propto 1/\sqrt{\Nk}$ for high temperatures (Fig.~\ref{fig_svext_pdLJ}),
$\svext(\Nk)$ should reveal an intermediate plateau (shoulder), $\Snonerg$, before it decays for even larger $\Nk$.  
A central question is then whether this intermediate plateau $\Snonerg(T)$ depends continuously on $T$
--- as suggested by our data (Fig.~\ref{fig_svext_pdLJ}) --- or if a jump-singularity appears \cite{Gardner}.
 
We have considered in the present work the standard deviations $\delta v$ associated with the empirical variance $v[\xbf]$, 
Eq.~(\ref{eq_vxdef}), with $p=2$. It is straightforward to generalize our approach to other moments $p$.
Especially, Eq.~(\ref{eq_key_1}) still holds and the generalized internal variance $\dvint$ must be given by 
a generalization of $\dvgauss[h]$, i.e. one expects
the same $V$-dependence for $h$ and $\svint \approx \svgauss$. 
Probing different moments $p$ should make manifest the higher-order spatial correlations
of the instantaneous stress field $\tauhat_{\rvec}$.
Note that the expectation values $v$ for $p=2,3,\ldots$ correspond to important contributions to the generalized 
stress-fluctuation formalism for the $p$-order elastic moduli $B_p$ (being the $p$-order strain derivative of the free energy) 
\cite{spmP1,Procaccia16}.
Surprisingly, the standard deviations $\delta B_p$ for $p > 2$ have been claimed to diverge
with increasing $V$ leading to a ``breakdown of nonlinear elasticity in amorphous solids" \cite{Procaccia16}. 
Since the common every day experience is rather that sufficiently large amorphous (plastic) bodies are 
well behaved according to standard continuum mechanics \cite{FerryBook,GraessleyBook,TadmorCMTBook},
the presented work suggests that the experimentally relevant standard deviations should be characterized 
by internal standard deviations $\sBpint$ using Eq.~(\ref{eq_dOint})
instead of the total standard deviations $\sBptot$ 
computed using Eq.~(\ref{eq_dOtot}) in Ref.~\cite{Procaccia16}.
We are currently working out the consequences of this idea.\footnote{The 
stress-fluctuation formalism for $B_p$ uses the fluctuations of stationary stochastic processes, i.e. no external
(linear) perturbation is applied to measure directly the moduli. It is unclear whether the out-of-equilibrium 
processes are described by the same fluctuations. It is an open theoretical question of how 
to generalize the fluctuation-dissipation relations, connecting the {\em average} linear out-of-equilibrium 
response to the {\em average} equilibrium relaxation \cite{DoiEdwardsBook,HansenBook,WKC16}, for their fluctuations.}

\section*{Author contribution statement}
JB and JPW designed the research project.
The presented theory was gathered from different sources by ANS and JPW.
GG (polymer films), LK (pLJ particles) and JPW (TSANET) performed the simulations and the data analysis.
JPW wrote the manuscript benefiting from contributions of all authors.

\section*{Acknowledgments}
We are indebted to O.~Benzerara for helpful discussions and acknowledge computational
resources from the HPC cluster of the University of Strasbourg.

\appendix

\section{System-size exponents $\gamint$ and $\gamext$}
\label{app_V}

We focus here on properties obtained for $\tsamp \gg \taubasin$.
The time dependence becomes thus irrelevant. 
Due to the non-ergodicity the $c$-dependence remains relevant, however,
and we compute $k$-averages $\Eop^{k}\ldots = \la \ldots \ra_c$ over all stochastic variables 
$x = \Eop^m x_m$ being themselves averages over $\Nm$ microscopic variables $x_m$
and compatible with the non-ergodicity constraint of the configuration $c$ considered.
Our task is to compute 
\begin{equation}
v = \Eop^c v_c \mbox{ and }
\Snonerg^2 = \Vop^cv_c \mbox{ for } v_c \equiv \la x^2 \ra_c - \la x \ra_c^2.
\label{eq_V_maintask}
\end{equation}
We assume that the microscopic variables $x_m$ are decorrelated as they come from uncorrelated microcells
and set $v_{cm} \equiv \la x_m^2 \ra_c - \la x_m \ra_c^2$ for the variance of the microscopic variable $x_m$.
Using the independence of the microcells $m$ yields 
\begin{eqnarray}
v_c & = & \frac{1}{\Nm} \times \left(\frac{1}{\Nm} \sum_m v_{cm} \right) \label{eq_V_1a} \\
\Vop^cv_c & = & \frac{1}{\Nm^3} \times \left( \frac{1}{\Nm} \sum_m \Vop^cv_{cm} \right) \label{eq_V_2a} 
\end{eqnarray}
where we have used that also the variances $v_{cm}$ are independent stochastic variables. 
Note that the $m-$averages (brackets) do not depend on $\Nm$ for large $\Nm$.
Hence,
\begin{equation}
v = \Eop^c v_c \propto 1/\Nm \mbox{ and } \Snonerg \propto 1/\Nm^{3/2}.
\label{eq_V_exponents}
\end{equation}
We have thus confirmed the exponents $\gaminthat \equiv \gamint+1=1$ and $\gamexthat \equiv \gamext + 1=3/2$ stated in Sec.~\ref{theo_V}
for uncorrelated microscopic variables.

\section{Distribution of $v_c$}
\label{app_px}

\begin{figure}[t]
\centerline{\resizebox{1.0\columnwidth}{!}{\includegraphics*{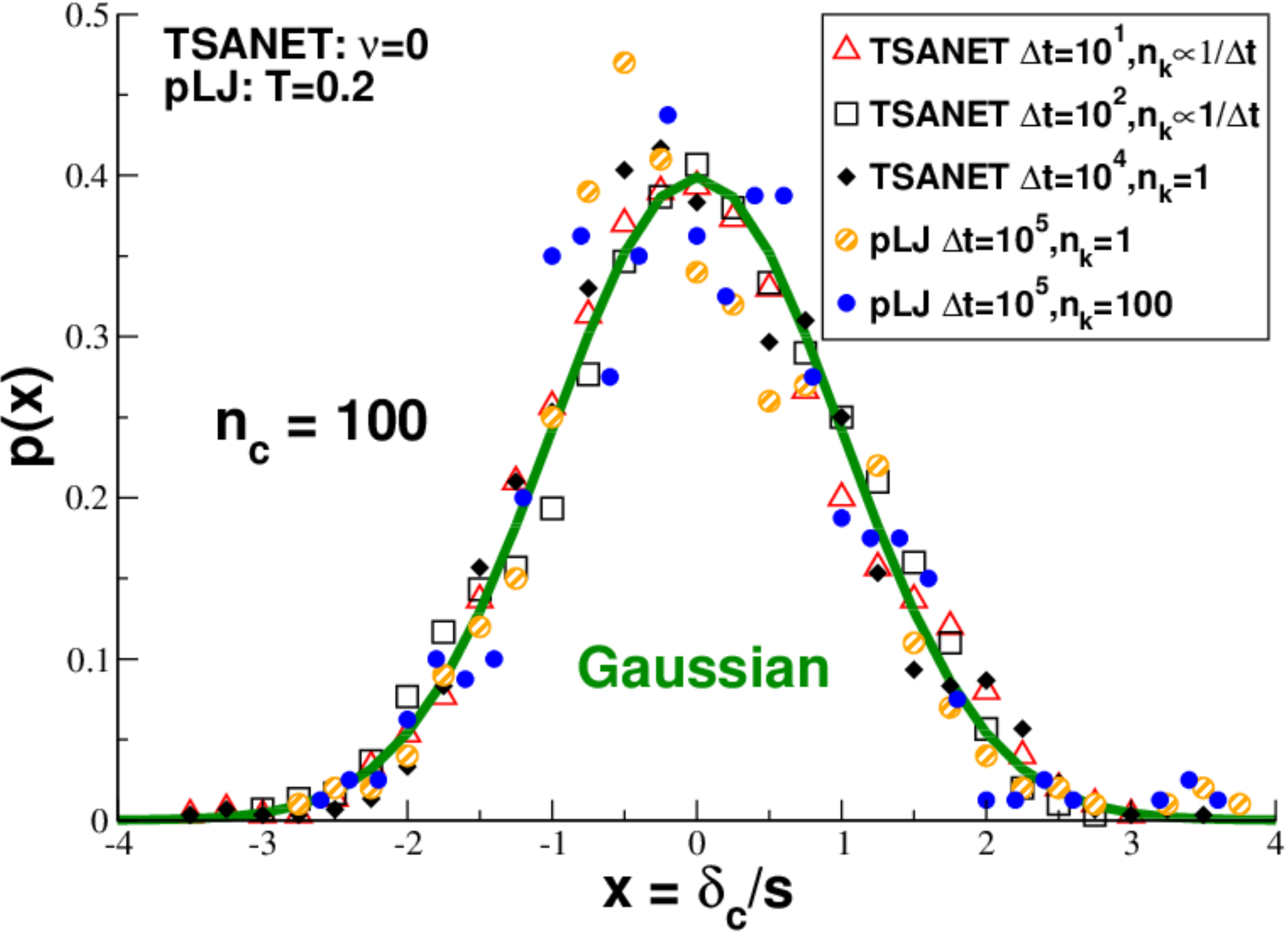}}}
\caption{Normalized histogram $p(x)$ for different $\tsamp$ and $\Nk$ as indicated.
The histograms are well described by a Gaussian (bold solid line).
}
\label{fig_px}
\end{figure}

Since $\dvext = \Vop^c v_c$ is finite, the $v_c = \Eop^k v[\xbf_{ck}]$ of different configurations $c$ must differ.
It is useful to rewrite Eq.~(\ref{eq_dOext}) by setting $v_c = v (1+\delta_c)$
in terms of the ``dimensionless dispersion" $\delta_c$. Using $\Eop^c \delta_c = 0$ we have
\begin{equation}
(\svext/v)^{2} \equiv s^2 = \ \Eop^c \delta_c^2 = \int \ddiff \delta_c \ p(\delta_c) \ \delta_c^2
\label{eq_dvext_b}
\end{equation}
with $s$ being the standard deviation of the normalized distribution $p(\delta_c)$.
For a Gaussian distribution all moments are set by $s$. 
In general, however, $p(\delta_c)$ may be non-Gaussian and may depend on the preparation history.
It may even happen in principle that some higher moments do not exist.
We present in Fig.~\ref{fig_px} the normalized distribution $p(x)$ for the 
rescaled dispersion $x = \delta_c/s$. A broad range of cases is considered.
The histograms are obtained using the $\Nc = 100$ independent configurations.
A reasonable data collapse on the Gaussian distribution (bold solid line) is observed.
This indicates that $\svext$ or $s$ are sufficient for the characterization of the distribution
of the dispersion $\delta_c$.
The Gaussianity was also checked by means of the standard non-Gaussianity parameter
\cite{HansenBook}, comparing the forth and the second moment of the distribution.
Clearly, an even larger number $\Nc$ is warranted in future work for a more critical test of 
the tails of the distribution using a half-logarithmic representation. 

\bibliographystyle{epj.bst}

\end{document}